\documentclass[showpacs,aps,amsmath,amssymb,twocolumn]{revtex4-2}

\usepackage[]{graphicx}
\usepackage{color}
\usepackage{bbold}
\usepackage[colorlinks=true, pdfstartview=FitV, linkcolor=blue, citecolor=red, urlcolor=black]{hyperref}
\newcommand{\be}{\begin{equation}}
\newcommand{\ee}{\end{equation}}
\newcommand{\ben}{\begin{eqnarray}}
\newcommand{\een}{\end{eqnarray}}
\newcommand{\bes}{\begin{subequations}}
\newcommand{\ees}{\end{subequations}}
\newcommand{\bb}{\bibitem}

\newcommand{\nn}{\nonumber\\}
\newcommand{\bfi}{\begin{figure}}
\newcommand{\efi}{\end{figure}}
\newcommand{\bc}{\begin{center}}
\newcommand{\ec}{\end{center}}
\newcommand{\sech}{\mbox{sech}}

\newcommand{\sn}{\mbox{sn}}
\newcommand{\dn}{\mbox{dn}}
\newcommand{\cn}{\mbox{cn}}

\begin{document}
\title{Impact of compactlike and asymmetric configurations of thick branes \\on the scalar-tensor representation of $f\left(R,T\right)$ gravity}
\author{Jo\~ao Lu\'is Rosa$^{1}$, A. S. Lob\~ao Jr.$^{2}$, and D. Bazeia$^{3}$}

\affiliation{$^1$Institute of Physics, University of Tartu, W. Ostwaldi 1, 50411 Tartu, Estonia\\
$^2$Escola T\'ecnica de Sa\'ude de Cajazeiras, Universidade Federal de Campina Grande, 58900-000 Cajazeiras, PB, Brazil\\
 $^3$Departamento de F\'isica, Universidade Federal da Para\'iba, 58051-970 Jo\~ao Pessoa, PB, Brazil}

\begin{abstract}
In this work we study braneworld models in generalized $f\left(R,T\right)$ gravity theories where the action depends on a function of the Ricci scalar $R$ and the trace of the stress-energy tensor $T$. We explore the so-called scalar-tensor representation of the theory, where the dependence on $R$ and $T$ are exchanged for two auxiliary real scalar fields. We introduce a first-order formalism to relate the auxiliary fields with the source field of brane and analyze four distinct possibilities of current interest regarding compactification and asymmetry of the brane. We investigate the behavior of the auxiliary fields and their stability in each one of the four cases. For the models where the solution of the source field engenders compact behavior, the auxiliary fields exhibit a hybrid structure, behaving differently inside and outside the compact region described by the source configuration. In particular, we found that the solutions of the auxiliary fields are significantly modified in models where the geometry is strongly influenced by the source field. Moreover, for the model capable of engendering asymmetric features, the two auxiliary fields also respond to modifications on the parameter that controls the asymmetry of the system.\\
\\
\end{abstract}

\maketitle{}
\vspace{1cm}

\section{Introduction}

Five-dimensional braneworld models are theories of gravity in a $(4,1)$ anti-de Sitter spacetime containing a four-brane in the presence of a single extra dimension of infinite extent. An interesting braneworld model was proposed by Randall and Sundrum (RS) in 1999; it is known as the RS model \cite{RS} and was motivated to provide an alternative explanation for the hierarchy problem. In a braneworld, gravity is supposed to flow through the extra dimension and the RS model is sometimes referred to as the thin brane model. Soon after, in Refs. \cite{T1,T2,T3} the thin brane scenario was modified to incorporate a real scalar field, the source field capable of generating a thick brane, in which the warp factor is now a smooth function of the extra dimension. In this thick brane scenario, the presence of the real scalar field may be seen as the source of the warped geometry that describes the five-dimensional braneworld.

The braneworld idea has been considered in several different contexts, for instance, to study the effects of a first-order phase transition \cite{C}, the presence of internal structure \cite{B1,B2,Dutra1} asymmetric scenarios  \cite{AS0,AS1,AS2,AS3,AS4,AS5,AS6,AS7,AS8} and other possibilities \cite{sken,review}. More recently, it has been studied to contextualize important questions concerning inflation, black holes, and dark energy, for instance. In particular, one notices that in Refs. \cite{Neves:2021dqx,Biggs:2021iqw,Banerjee:2021qei} the authors examine black hole dynamics in braneworld scenarios. Most of the issues considered are based on proposals where the braneworld scenario is generalized by including new terms in the Einstein-Hilbert action \cite{Davood:2018,Sui:2020fty}. Consequently, generalized functions of Lorentz invariant terms have been included in the standard action, to find stable solutions in brane models with $F(R)-$gravity \cite{Rippl:1995bg, Hwang:2001pu,DeFelice:2010aj,RMP,Bazeia:2013oha,Bazeia:2013uva,Olmo,Olmo01}, Gauss-Bonnet \cite{GB01,GB02,Bazeia:2015dna}, trace $T$ of the energy-momentum tensor \cite{Bazeia:2015owa}, and other extended theories of gravity. In the context of generalized models, a specific issue concerns the emergence of inner structures of the brane. When this is the case, studies indicate that the internal modification of the brane may change the resonance spectrum and location of the graviton when the brane is coupled with fermions \cite{Liu1,Gomes1,Dutra2,Moreira:2021vcf,Yang:2012hu,Yu:2015wma,Castro}. This is interesting, and gravitational resonances have also been studied in teleparallel gravity, which may be considered as a possible candidate for dark matter \cite{Tan:2020sys}.

An interesting way to analyze generalized gravity models is through the scalar-tensor representation. In this scenario, auxiliary scalar fields are introduced to replace invariant contributions added to the model, but yet carrying all the information encoded within. In this representation, we can describe the dynamics of auxiliary fields separately and, if the brane has an inner structure, it will be determined in the model by the solutions of the auxiliary fields. However, due to the complexity of the equations of motion, the solutions of the auxiliary fields are in general obtained via numerical methods, using as ansatz the solution of the source field of the brane in models that develop analytical solutions. In previous works, for instance, in Refs. \cite{Rosa:2021tei, Rosa:2021teg}, it was demonstrated that, in the scenario where the brane is generalized by the inclusion of a general function of the Ricci scalar and the trace of the stress-energy in the form $F(R,T)$ \cite{Lobo1}, the fitting parameters imposed on the ansatz of the source field are less significant than the boundary conditions used to solve the equations of motion that come from the scalar-tensor representation. In order to further explore this behavior, in Ref. \cite{Rosa:2021ld}, the dynamics of the source field was changed by including a cuscuton term. In this new scenario, we verified a quantitative change in the behavior of the auxiliary fields, with the cuscuton term having a non-negligible influence on the inner structure of the zero-mode which appears from the stability of the gravitational sector.

The modification of gravitational theories by the inclusion of non-standard kinetics, such as cuscuton, has been the subject of interesting investigations in recent years, in particular as they offer new possibilities for solving important cosmological questions such as dark energy and dark matter. In the braneworld scenario, models with K-fields were first presented in \cite{Adam:2007ag} and also generalized in \cite{Bazeia:2008zx,Bazeia:2008tj}. On the other hand, the cuscuton dynamics, which has been studied in Refs. \cite{C01,C02,C03} to find exact solution describing an accelerating four-dimensional universe with a stable extra dimension \cite{C01}, the possibility to construct analytic solutions for the cosmological background evolution that mimics $\Lambda$CDM cosmology \cite{C02}, and the
cuscuton gravity as a classically stable limiting curvature theory \cite{C03}, the cuscuton dynamics has also been very recently investigated within the braneworld context in Refs.  \cite{Andrade:2018afh,Bazeia:2021jok}. Due to this, it is of current interest to understand if other modifications in the structure of the source field of the brane, such as the assumption that they describe compact structures, can also affect the behavior of the auxiliary fields. In this respect, we know that the compact structure of the source field may induce a new behavior on the model, producing what is known as hybrid brane, i.e., a brane that behaves differently, having a thin or a thick profile inside or outside the compact space \cite{Bazeia:2014hja}. Moreover, we can also induce a compression of the brane geometry, as is done in Ref. \cite{Bazeia:2015eta}, to check how the auxiliary fields are modified in these cases. It is also of interest to see how the auxiliary fields modify when one changes the thick braneworld scenario to admit asymmetric configurations.

To examine the above issues, we organize the present work as follows. In Sec. \ref{formalismo} we introduce the general formalism describing a $f(R,T)-$brane in the tensor-scalar representation, and describe the first-order formalism to explain the interdependence between the auxiliary scalar fields and the solution of the source field. In Sec. \ref{models}, we study several models, two of them to understand how the appearance of compactlike configurations changes the behavior of the auxiliary fields, and another one, to describe effects of the asymmetry of the brane. In the first model, we introduce a simpler possibility as a warm-up for the following investigations. In particular, in the second model, we consider solutions with a compactlike profile for the source field of the brane to study the changes in the auxiliary fields. Moreover, in the third model we also study modifications in the geometry of the brane, to see how it affects the auxiliary fields. We also study another possibility, the fourth model which allows for the presence of asymmetric configurations. There we explore how the warp factor, the auxiliary fields $\varphi$ and $\psi$ and the potential $U$ modify as we change the asymmetric profile of the system. In Sec. \ref{secStability} we investigate linear stability of the gravitational sector, paying closer attention to the presence of the compactlike behavior and the asymmetry of the model. We end the work in Sec. \ref{sec:conclusion}, including our final comments and conclusions.

\section{Formalism}\label{formalismo}

In this section we describe the scalar-tensor representation to study generalized braneworld models of the $f\left(R,T\right)$ type, where $R$ is the Ricci scalar, and $T$ is the trace of the stress-energy tensor $T_{ab}$. For this purpose, we assume the standard metric used in the braneworld scenario with an additional dimension $y$ described by a line element given by
\begin{equation}\label{metricbrane}
    ds^2=e^{2A(y)}\eta_{\mu\nu}dx^\mu dx^\nu-dy^2\,,
\end{equation}
where $\eta_{\mu\nu}$ is the four-dimensional Minkowiki metric with signature $(+\,-\,-\,-)$, written in terms of a set of coordinates $x^\mu$ and $A(y)$ is the warp function, which is supposed to depend only on the extra dimension. In this case, the Ricci scalar can be written as $R=8A''+20A'^2$, where the prime denotes derivative with respect to the extra dimension. Also, Greek indices $\mu,\nu, ...$ run from $0$ to $3$ and Latin indices $a,b, ...$ run from $0$ to $4$.

Let us now follow Ref. \cite{Bazeia:2015owa} and consider a generalized action $S$ in five-dimensions coupled to a scalar source field as
\begin{equation}\label{actiongeo}
S=\frac{1}{2\kappa^2}\int_\Omega\!\!\sqrt{|g|}\,d^5x\Big[f\left(R,T\right)-2\kappa^2{\cal L}_s\Big],
\end{equation}
where $\kappa^2=8\pi G_5/c^4$ is a coupling constant, with $G_5$ the 5-dimensional gravitational constant, and $c$ the speed of light, $\Omega$ is a five-dimensional spacetime manifold described by a set of coordinates $x^a$ and $|g|$ is the absolute value of the determinant of the metric $g_{ab}$. In this work, we consider that the Lagrangian density ${\cal L}_s$ has the form
\begin{equation}\label{lagrangeStandar}
    {\cal L}_s=\frac12\nabla_a\phi\nabla^a\phi-V(\phi),
\end{equation}
where $\phi$ is a source field of the brane and $\nabla_a$ represents the covariant derivatives.

We can also obtain the stress-energy tensor $T_{ab}$, defined in the usual way as the variation of the Lagrangian density  with respect to the metric $g_{ab}$. Using Eq. \eqref{lagrangeStandar} we obtain,
\begin{equation}\label{defTab}
    T_{ab}=\,\nabla_a\phi\nabla_b\phi-g_{ab}\,\mathcal L_s.
\end{equation}
To obtain the trace of the stress-energy tensor, we make the contraction with the metric, i.e.,  $g^{ab}T_{ab}$, thus,
\begin{equation}\label{defTraço}
    T=-\frac32\nabla_a\phi\nabla^a\phi+5V(\phi).
\end{equation}

We could now obtain the equations of motion for the source field and the modified field equations by varying Eq. \eqref{actiongeo} directly with respect to $\phi$ and $g_{ab}$, respectively. However, let us perform a transformation to a dynamically equivalent scalar-tensor representation by introducing two new auxiliary scalar fields $\varphi$ and $\psi$ instead. As it was discussed in Ref. \cite{Rosa:2021tei}, if $f_{RR}f_{TT}\neq f_{ RT }^2$, where the subscripts $R$ and $T$ denote partial derivatives of the function $f$ with respect to these variables, respectively, we can express Eq. \eqref{actiongeo} as
\ben
S=\frac{1}{2\kappa^2}\!\!\int_\Omega\!\!\!\sqrt{|g|}\,\Big(\varphi R+\psi T-U\left(\varphi,\psi\right)
-2\kappa^2{\cal L}_s\Big)d^5x,\nn\label{actionst}
\een
where the scalar fields $\varphi$ and $\psi$ and the scalar interaction potential $U\left(\varphi,\psi\right)$ are defined in terms of the function $f\left(R,T\right)$ as
\ben
\varphi=\frac{\partial f}{\partial R}, \quad \psi=\frac{\partial f}{\partial T},\quad U=\varphi R+\psi T-f.
\een
Now we can proceed to obtain the equations of motion by treating each of these new fields as an independent quantity. Note that Eq. \eqref{actionst} now depends on four quantities: the metric $g_{ab}$, the auxiliary scalar fields $\varphi$ and $\psi$, and the source field of the brane $\phi$. Varying Eq.~\eqref{actionst} with respect to the auxiliary scalar fields, $\varphi$ and $\psi$, one obtains
\begin{equation}\label{eomphi}
    U_\varphi=R,\qquad\qquad U_\psi=T,
\end{equation}
where $U_\varphi=\partial U/\partial\varphi$ and $U_\psi=\partial U/\partial\psi$. The modified field equations obtained from a variation of Eq. \eqref{actionst} with respect to $g_{ab}$ have the form,
\begin{eqnarray}
    &&-\frac{1}{2} g_{ab}\left(\varphi R-U\right)  +\varphi R_{ab}-\left(\nabla_a\nabla_b-g_{ab}\,\Box\right)\varphi
    \nn
    &&\qquad   =\kappa^2 T_{ab}+\frac32\psi\nabla_a\phi\nabla_b\phi +\frac12\,g_{ab}\,\psi T ,
    \label{fieldst}
\end{eqnarray}
where $\Box\equiv \nabla_c\nabla^c$ is the d'Alembert operator and $R_{ab}$ is the Ricci tensor. Finally, the equation of motion for the source field $\phi$ is given by,
\ben\label{eq142}
\nabla_a\nabla^a\phi+V_\phi =-\frac{3}{2\kappa^2}\nabla_a\big(\psi \nabla^a\phi\big)-\frac{5}{2\kappa^2}\psi V_\phi,
\een
where $V_\phi=dV/d\phi$.

As usual, we now assume that the system is static and that all scalar fields are functions of the extra dimension only, i.e. $\psi=\psi(y)$, $\varphi=\varphi(y)$ and $\phi=\phi(y)$. Moreover, in the following calculations we will consider a system of geometric units for which $\kappa^2=2$ holds. Using this prescription, the trace of the stress-energy tensor can be written as $T=3\phi^{\prime 2}/2+5V$. We then obtain the equations of motion for the auxiliary fields as
\begin{equation}\label{kgphi}
    U_\varphi=8A''+20A'^2,
\end{equation}
\begin{equation}\label{kgpsi}
    U_\psi=\frac32\phi^{\prime 2}+5V.
\end{equation}
On the other hand, given the isotropy of the system in the four-dimensional spacetime, Eq. \eqref{fieldst} features only two non-vanishing and linearly independent components, which are
\ben
&&6\varphi\big( A''+2A'^2\big)+6\varphi' A'+2\varphi''-U=\nn
&&\qquad\qquad  =-2\phi^{\prime 2}-4V-\big(\frac32\phi^{\prime 2}+5V\big)\psi,\label{field1}\\
&&-12\varphi A'^2-8\varphi' A'+U=-2\phi^{\prime 2}+4V-\nn
&&\qquad\qquad-\big(\frac32\phi^{\prime 2}-5V\big)\psi.\label{field2}
\een
Furthermore, the equation of motion for the source field $\phi$ given by Eq. \eqref{eq142} yields
\begin{eqnarray}
\phi''\!+4A'\phi'=V_\phi-\frac{3}{4}\phi'\psi'\!-\big(\frac{3}{4}\phi''\!+3A'\phi'\!-\frac{5}{4}V_\phi\big)\psi.\qquad\label{kgchi}
\end{eqnarray}

It is possible to show that from the five equations given by Eqs. \eqref{kgphi} to \eqref{kgchi}, only four are linearly independent. To prove this statement, one takes the derivative of Eq.~\eqref{field2} with respect to $y$, uses Eq. \eqref{field1} to eliminate the dependency on $A''$, and uses Eqs. \eqref{kgphi}, \eqref{kgpsi}, and \eqref{kgchi} to eliminate the dependencies in $U_\varphi$, $U_\psi$, and $\phi''$, respectively. As a result, one recovers Eq. \eqref{field2} itself, thus proving that this equation is linearly dependent on the others. Consequently, one can simplify the system by exchanging the two field equations in Eqs. \eqref{field1} and \eqref{field2} by a linear combination of themselves to obtain the simpler relation
\begin{equation}\label{fieldsum}
    3\varphi A''+\varphi''-\varphi'A'=-2\phi'^2-\frac32\psi\phi^{\prime 2}.
\end{equation}
Moreover, it was show in Ref. \cite{Rosa:2021tei} that $U_\varphi$, $U_\psi$ and $U$ can all be regarded as independent quantities due to the presence of two degrees of freedom in $U$, coming from the two arbitrary dependencies of this function in the scalar fields $\varphi$ and $\psi$. Therefore, we can use the chain rule to write the potential $U$ as $U'=U_\varphi \varphi'+U_\psi\psi'$. If we now consider the Eq. \eqref{kgphi} and Eq. \eqref{kgpsi}, we get
\begin{equation}\label{Urelfinal}
U'=\big(8A''+20A'^2\big)\varphi'+\big(\frac{3}{2}\phi'^2+5V\big)\psi'\,.
\end{equation}
This equation already encodes all the information from Eqs. \eqref{kgphi} and \eqref{kgpsi}, while reducing the number of degrees of freedom in $U$ from two (explicit dependencies in $\varphi$ and $\psi$) to one (dependency on $y$ only). Thus, Eq. \eqref{Urelfinal} replaces Eqs. \eqref{kgphi} and \eqref{kgpsi}. We are thus left with a simpler system of three equations, i.e., Eqs. \eqref{kgchi}, \eqref{fieldsum} and \eqref{Urelfinal}, for the six independent degrees of freedom carried by the quantities $\varphi$, $\psi$, $\phi$, $U$, $V$, and $A$. This system is thus under-determined and one needs to impose three extra constraints to close the system. Let us proceed considering the first-order formalism in which we rewrite the functions $A$, $\phi$, and $V$ in terms of a known function $W$ that depends solely on the source field $\phi$, i.e.,
\begin{equation}\label{FOF}
    \phi^{\prime}=W_\phi\,\,\,,\qquad A^{\prime}=-\frac23W\,,
\end{equation}
 where $W_\phi=dW/d\phi$. Note that if we know the function $W(\phi)$, the first-order formalism must restrict the form of the solution of the source field and the warp function $A$. In this paper, we are interested in studying the behavior of the auxiliary fields $\varphi$ and $\psi$ when the source field is a topological defect solution. To guarantee the asymptotic behavior of the solution of the source field, we will assume that the potential has the standard form
\begin{equation}\label{PotV}
    V(\phi)=\frac12W_\phi^2-\frac43W^2\,.
\end{equation}
This is an important choice and it allows that we rewrite the energy density as a total derivative in the form,
\begin{equation}\nonumber
    \rho(y)=\frac{d}{dy}\big(e^{2A}W\big).
\end{equation}

At this point we have already introduced two constraints on the system, i.e., Eqs. \eqref{FOF} and \eqref{PotV} (note that the two equations in Eq. \eqref{FOF} contribute with a single constraint to the system since an extra degree of freedom was added in the quantity $W$). The last constraint imposed on the system will be the explicit form of the function $W$, which will be introduced in the following sections. Introducing the constraints from Eqs. \eqref{FOF} and \eqref{PotV} into Eq. \eqref{kgchi} one obtains a connection equation for the field $\psi$ in the form,
\begin{eqnarray}\label{eqpsi}
9\psi'+2\left(8W-3W_{\phi\phi}\right)\psi=0.
\end{eqnarray}
Although we cannot solve this equation in general, we will show that it is possible to obtain analytic solutions for some specific situations. The two remaining equations in the system, i.e., Eqs. \eqref{fieldsum} and \eqref{Urelfinal}, under these assumptions take the forms
\begin{equation}\label{eqvarphi}
    -2\varphi W_\phi^2+\varphi''+\frac23\varphi'W+2W_\phi^2+\frac32\psi W_\phi^2=0,
\end{equation}
and
\begin{equation}\label{eqU}
    U'=4\Big(\frac{5}3W^2-W_{\phi}^2\Big)\Big(\frac{4}3\varphi'-\psi'\Big)\,.
\end{equation}

To demonstrate how the formalism presented in this Section works, we will consider some specific models that produce interesting braneworld scenarios which are geometrically stable against first-order perturbations of the metric.

\section{Specific Models}\label{models}

Let us now proceed to some applications of the scalar-tensor formalism via the first order procedure described in the previous Section. In this sense, we will start by defining each distinct model via the explicit form of the function $W=W(\phi)$, which depends solely on the source scalar field $\phi$.

\subsection{First model}\label{modeloA}

\begin{figure}[!htb]
    \begin{center}
        \includegraphics[scale=0.6]{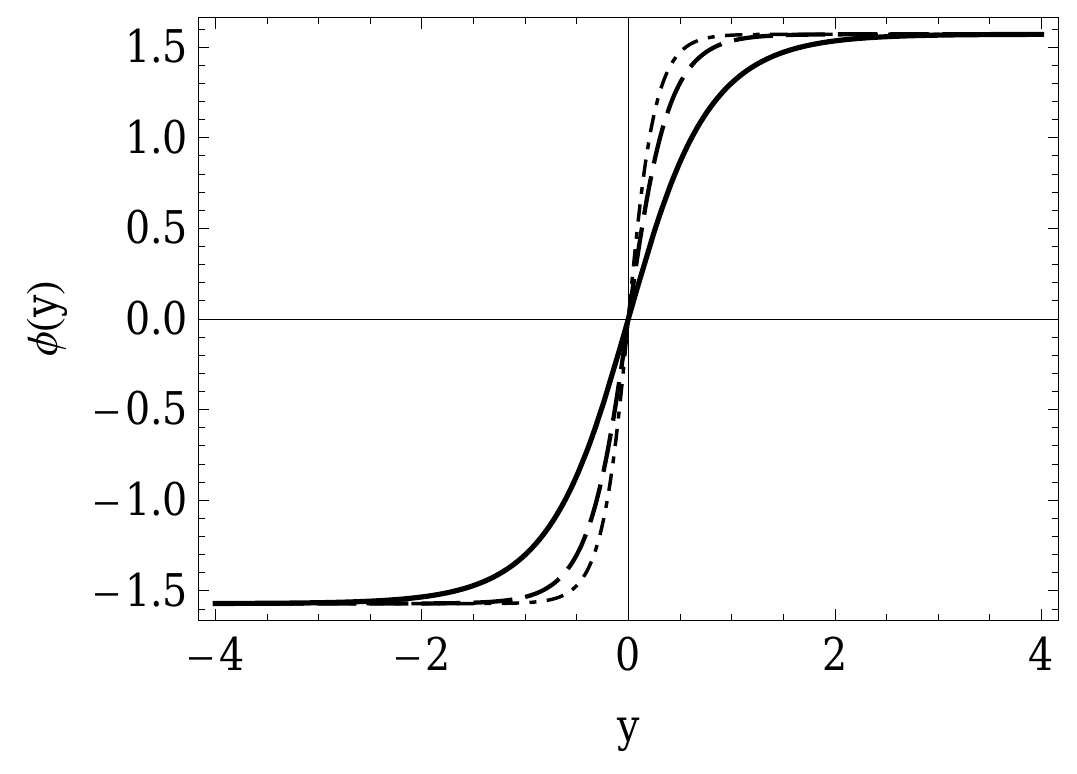}
    \end{center}
    \vspace{-0.9cm}
    \caption{\small{Kink solution given by Eq. \eqref{SSG} for $a=1$ (solid line), $a=2$ (dashed line) and $a=3$ (dot-dashed line)}. \label{fig1}}
\end{figure}

As a first case, let us consider that the $W(\phi)$ function is described by the sine-Gordon model, i.e.,
\begin{equation}\label{WSG}
    W(\phi)=2a\sin\phi,
\end{equation}
where $a$ is a real parameter. For $a>0$ we have kink solutions and for $a<0$ we have anti-kink solutions, which essentially present the same physics, and thus in the following we shall deal with $a>0$ without loss of generality. We know that this model generates stable solutions in situations with standard gravity. Thus, it is a good starting point to analyze the changes induced by the introduction of the auxiliary functions described in the previous section. Considering the first of Eqs.~\eqref{FOF}, we obtain a kink-like solution for $\phi$ in the form,
\begin{equation}\label{SSG}
    \phi(y)=2\arctan\big[\tanh(ay)\,\big],
\end{equation}
where the parameter $a$ controls the characteristic width of the solution, without changing its asymptotic value $\phi_v=\pm\pi/2$ when $y\to\pm\infty$ as we can observe in Fig.~\ref{fig1}, for the values $a=1, 2, 3$. Note that as $a$ increases, the source field solution becomes more and more concentrated near the origin.

In the limit where $a$ goes to infinity, the solution behaves like a sign function, defined by
\begin{eqnarray}\nonumber
   \mbox{sign}(y) = \left\{
    \begin{array}{clc}
        -1 &  \mbox{ to } &y<0\,,\\
        0  &  \mbox{ to } &y=0\,,\\
         1 & \mbox{ to }  &y>0\,.
    \end{array} \right.
\end{eqnarray}

The potential $V(\phi)$ obtained using Eq. \eqref{WSG} in Eq.~\eqref{PotV} can be written as
\begin{equation}
    V(\phi)=2 a^2-\frac{22}{3} a^2 \sin^2(\phi).
\end{equation}
For the asymptotic value of the solution, the potential assumes the value $V(\phi_v)= -16a^2/3$. In this limit, $V$ plays the role of a cosmological constant $\Lambda_5\equiv V(\phi_v)$, and thus as $\Lambda_5<0$ the Bulk is asymptotically $AdS_5$.

\begin{figure}[!htb]
    \begin{center}
        \includegraphics[scale=0.6]{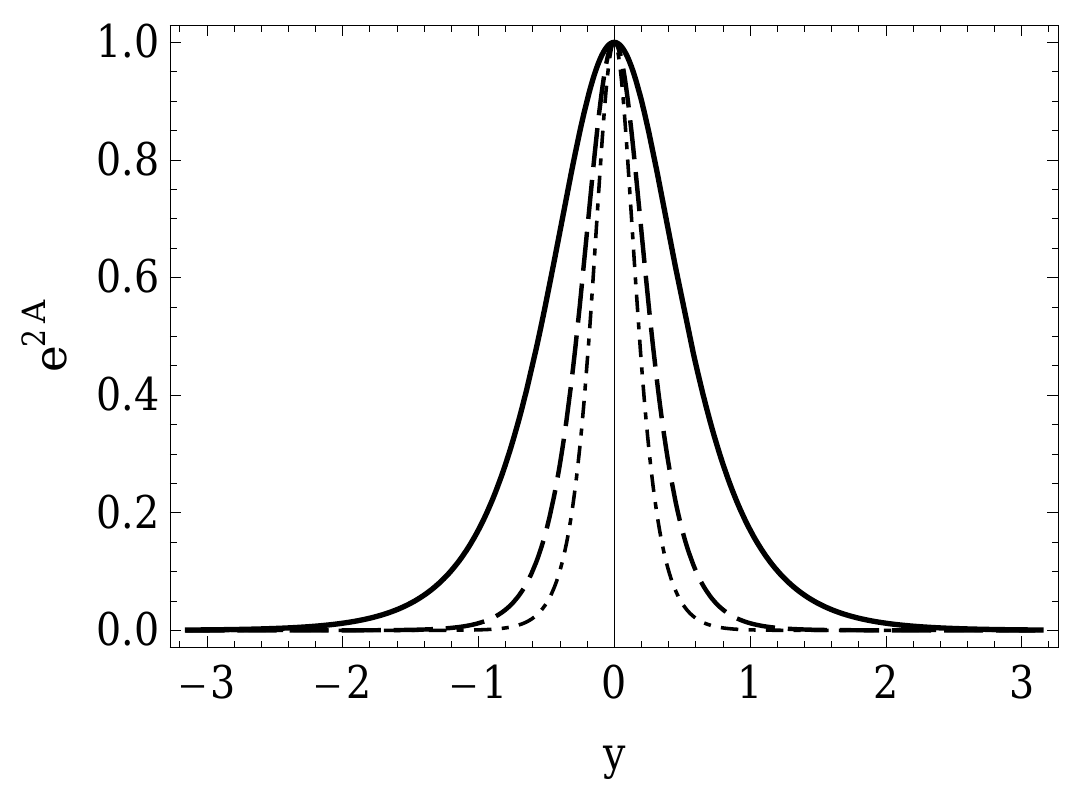}
    \end{center}
    \vspace{-0.9cm}
    \caption{\small{Warp factor $e^{2A}$ given by Eq. \eqref{IIIAwarp} for $a=1$ (solid line), $a=2$ (dashed line) and $a=3$ (dot-dashed line)}. \label{fig2}}
\end{figure}

The warp function is obtain from the second of Eqs.~\eqref{FOF}, using the Eq. \eqref{WSG} and the solution in Eq.~\eqref{SSG}, and it takes the form
\ben\label{IIIAwarp}
A(y)=(2/3)\ln\big[\sech(2a y)\big].
\een
In Fig. \ref{fig2} we show the warp factor $e^{2A}$  for the values $a=1, \,2,\, 3$. Note that just like the source field solution, the warp factor also becomes more and more concentrated as the $a$ parameter increases.

We verified that the Kretschmann scalar defined by
\begin{equation}\nonumber
    K=40A^{\prime 4}+16A^{\prime\prime 2}+32A^{\prime 2}A^{\prime\prime}\,,
\end{equation}
exhibits the expected behavior, i.e., it remains finite for the entire range of the extra dimension $y$. In particular, $K(0)=(32a/3)^2$ and $ K(y\to\pm\infty)= 40(4a/3)^4$.

\begin{figure}[!htb]
    \begin{center}
        \includegraphics[scale=0.6]{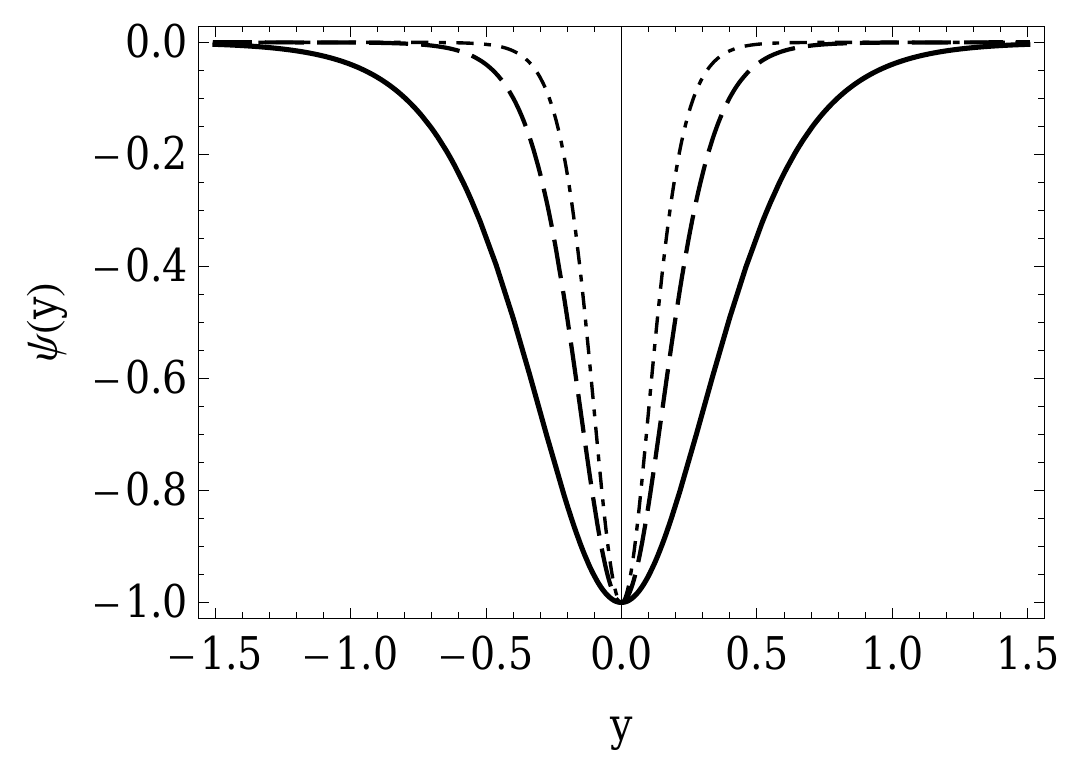}
    \end{center}
    \vspace{-0.9cm}
    \caption{\small{Field $\psi(y)$ given by Eq. \eqref{Solpsi} for $\psi_0=-1$, $a=1$ (solid line), $a=2$ (dashed line) and $a=3$ (dot-dashed line)}. \label{fig3}}
\end{figure}

Let us now analyze the behavior of auxiliary fields starting with $\psi(y)$. For this, we use the Eq. \eqref{WSG} and the solution in Eq. \eqref{SSG} to write Eq. \eqref{eqpsi} as
\begin{eqnarray}
9 \psi'+44 a \tanh(2 a y) \psi=0\,.
\end{eqnarray}
This differential equation  can be solved analytically and yields solutions of the form,
\begin{equation}\label{Solpsi}
    \psi(y)=\psi_0\, \sech^{22/9}(2ay)\,,
\end{equation}
where $\psi_0$ is an integration constant. Fig. \ref{fig3} show the behavior of field $\psi$ given by Eq. \eqref{Solpsi}. In this plot we used $\psi_0=-1$ and $a=1,\, 2,\, 3$. As expected, this field behaves similarly to the warp factor close to the origin. This result suggests that this field also has a strong tendency to shrink around $y=0$, displaying compacticity. We have verified that for other negative values of $\psi_0$ the general behavior of the field does not change qualitatively. However, if $\psi_0$ becomes positive, then the corresponding solution is reflected around the $y$ axis, exhibiting a bell-shaped behavior. In any case, the field always vanishes asymptotically.

\begin{figure}[!htb]
    \begin{center}
        \includegraphics[scale=0.6]{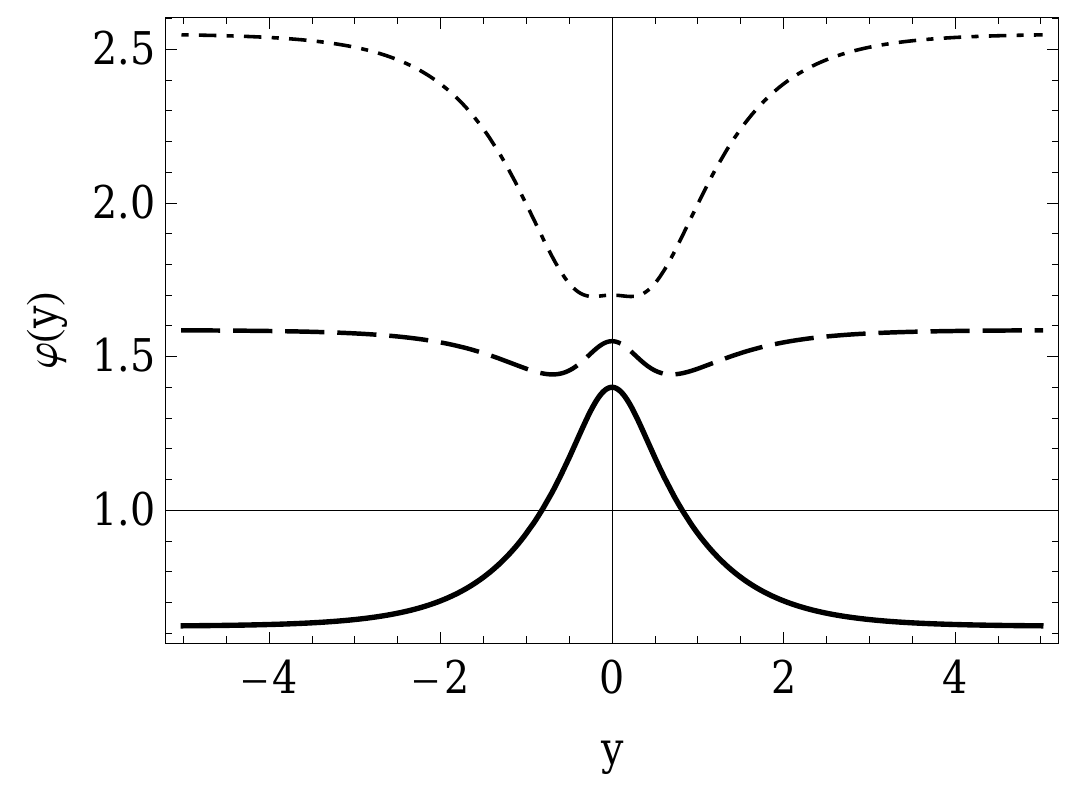}
    \end{center}
    \vspace{-0.9cm}
    \caption{\small{Plot of field $\varphi(y)$ for $a=1$, $U_0=0$, and $\psi_0=1$ with $\varphi_0=1.4$ (solid line), $\varphi_0=1.55$ (dashed line), and $\varphi_0=1.7$ (dotdashed line)}. \label{fig4}}
\end{figure}

\begin{figure}[!htb]
    \begin{center}
        \includegraphics[scale=0.6]{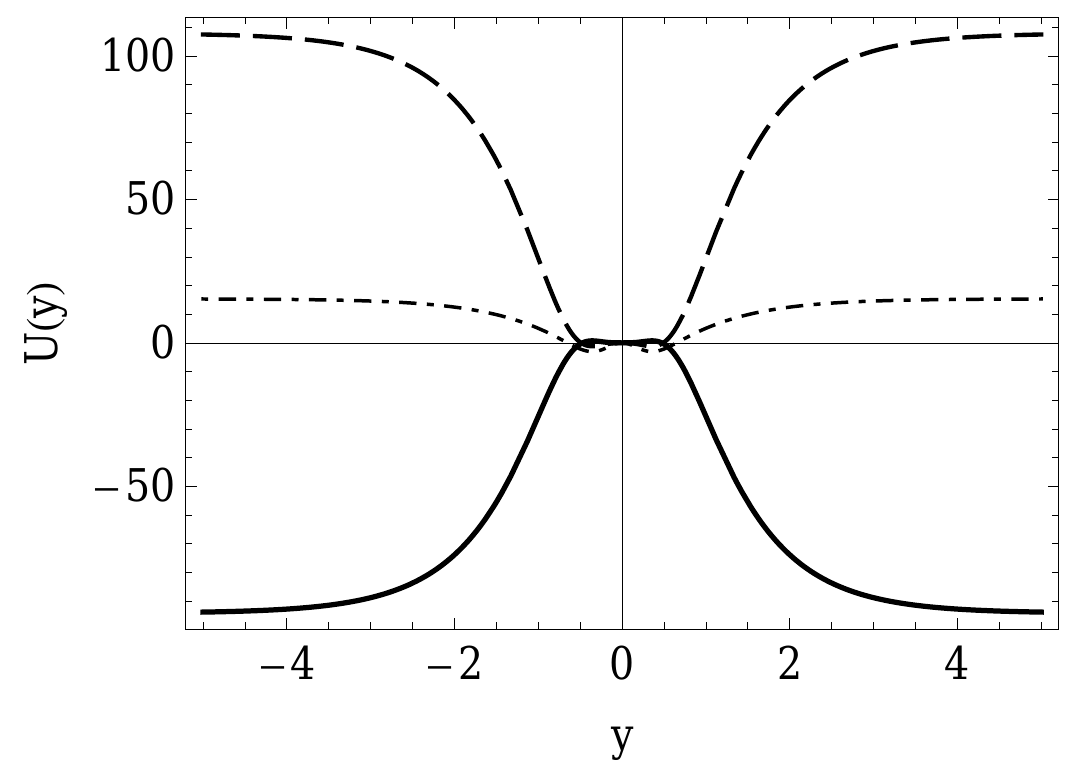}
        \end{center}
    \caption{\small{Plot of potential $U(y)$ for $a=1$, $U_0=0$ and $\psi_0=1$  with $\varphi_0=0.8$ (solid line), $\varphi_0=1.3$ (dashed line), and $\varphi_0=1.6$ (dotdashed line)}. \label{fig5}}
\end{figure}

The other two quantities, i.e., the fields $\varphi(y)$ and the potential $U(y)$ can only be obtained through numerical integrations of Eqs. \eqref{eqvarphi} and Eq. \eqref{eqU}, subjected to the boundary conditions $\varphi(0)=\varphi_0$, $\varphi'(0)=0$ and $U(0)=U_0$, for some constant parameters $\varphi_0$ and $U_0$. Furthermore, since these equations also feature an explicit dependency in $\psi$, the results will also be influenced by the parameter $\psi_0$. Depending on the values of $\varphi_0$ and $\psi_0$, the solution for $\varphi$ can have a plethora of different qualitative behaviors. For example, taking $\psi_0=1$, one verifies that for $\varphi_0\gtrsim 1.75$ the solution for $\varphi$ presents a single global minimum at $y=0$ and grows outwards, for $\varphi_0\lesssim 1.5$ the solution for $\varphi$ presents a single global maximum at $y=0$ and decreases outwards, and for $1.5\lesssim\varphi_0\lesssim1.75$ the field $\varphi$ undergoes a phase transition between these two behaviors. These results are show in Fig. \ref{fig4}. As for the solution for $U$, we verified that the parameter $U_0$ functions solely as a translation along the vertical axis, and thus we have considered $U_0=0$ for simplicity, without loss of generality. The shape of $U$ is also qualitatively influenced by the parameters $\varphi_0$ and $\psi_0$. Taking $\psi_0=1$, one verified that $U$ presents a local minimum at $y=0$ surrounded by two maxima and an outwards decrease for $\varphi_0\lesssim 0.8$, a local maximum at $y=0$ surrounded by two minima and an outwards increase for $\varphi_0\gtrsim 1.5$, and a phase transition between these two behaviors for $0.8\lesssim\varphi_0\lesssim1.5$. These results are shown in Fig. \ref{fig5}. For both $\varphi$ and $U$, the effects of the parameter $a$ are the same as before, i.e., to contract the solutions closer to $y=0$, and thus we chose not to include an analysis of this parameter in the figures.

\subsection{Second model}\label{modeloB}

\begin{figure}[!htb]
    \begin{center}
        \includegraphics[scale=0.6]{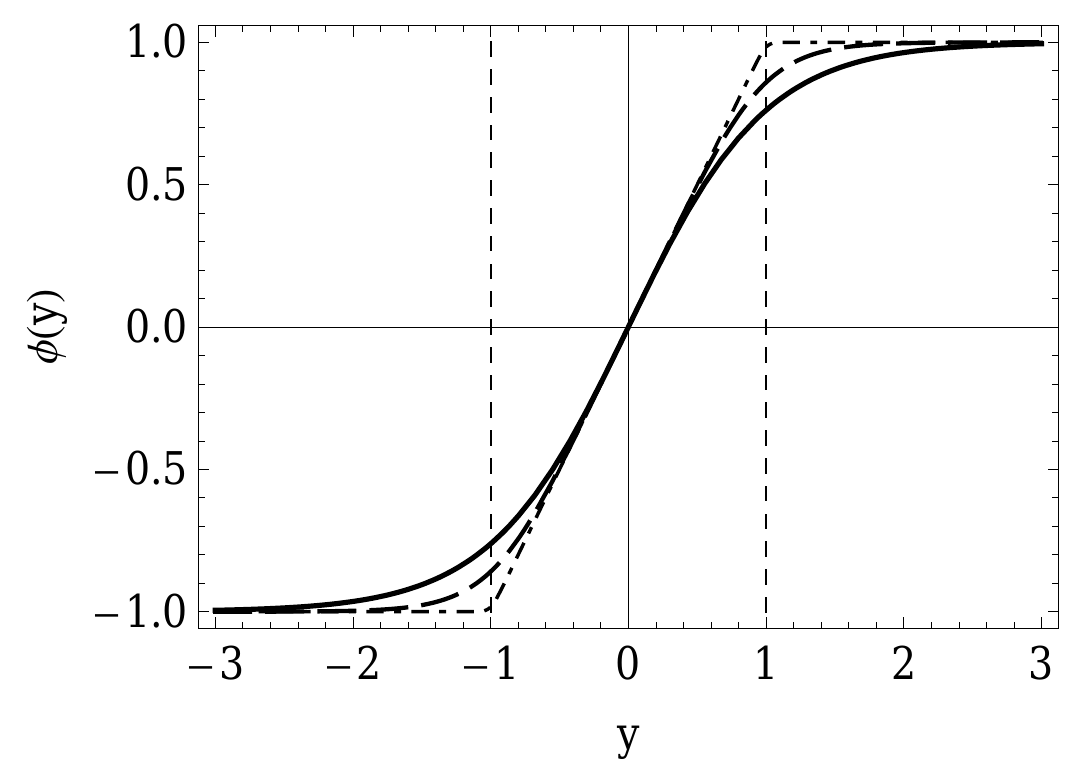}
    \end{center}
    \vspace{-0.9cm}
    \caption{\small{Kink solution obtain by Eq. \eqref{eqphicomp} for $n=1$ (solid line), $n=2$ (dashed line) and $n=20$ (dot-dashed line).} \label{fig6}}
\end{figure}

Let us now analyze the effects of a compactlike behavior of the source field $\phi$ in the auxiliary fields $\psi$, $\varphi$, and the potential $U$. As it was shown in Ref. \cite{Bazeia:2014hja}, we can obtain compact kinks in models with standard dynamics if we consider $W(\phi)$ in the form
\begin{equation}\label{WComp}
    W(\phi)=\phi-\frac{\phi^{2n+1}}{2n+1},
\end{equation}
with $n$ a positive integer, i.e., $n\geq 1$. Note that for $n=1$, we recover to the $\phi^4$ model. In Ref. \cite{Rosa:2021tei} the authors studied brane models in the scalar-tensor representation having kinklike solutions for the source field in the $\phi^4$ model. While in the previous work the solutions were free to live in every region of the space, here the solutions may shrink towards a compact space as $n$ increases to larger values. To verify this, we can take Eq. \eqref{WComp} and the first of Eqs. \eqref{FOF} to write
\begin{equation}\label{EqPOComp}
    \phi'=1-\phi^{2n}\,.
\end{equation}
The general solution of this equation can be written in terms of the hypergeometric function ${}_2F_{1}$ in the form,
\begin{equation}\label{eqphicomp}
  \phi\times{}_2F_{1}\Big(\frac1{2n},\, 1,\, \frac{2n+1}{2n},\phi^{2n}\Big)=y\,.
\end{equation}
Unfortunately, this equation is not invertible in general for any value of $n$, and thus we plot the results numerically in Fig. \ref{fig6}. We can see that when $n$ grows the solution becomes more and more concentrated in the compact region between $y=\pm 1$. Close to the origin, the solution behaves as a straight line. This can be seen if we expand Eq. \eqref{eqphicomp} in a Taylor series around the origin and take the large $n$ limit, where we obtain $\phi(y)\approx y$. Thus, in the limit at which the solution becomes compact, we have
\begin{eqnarray}\nonumber
    \phi(y) = \left\{
    \begin{array}{clc}
        y  &  \mbox{ for } &|y|\leq 1\,,\\
        \mbox{sign}(y) & \mbox{ for }  &|y|>1\,.
    \end{array} \right.
\end{eqnarray}

\begin{figure}[!htb]
    \begin{center}
        \includegraphics[scale=0.6]{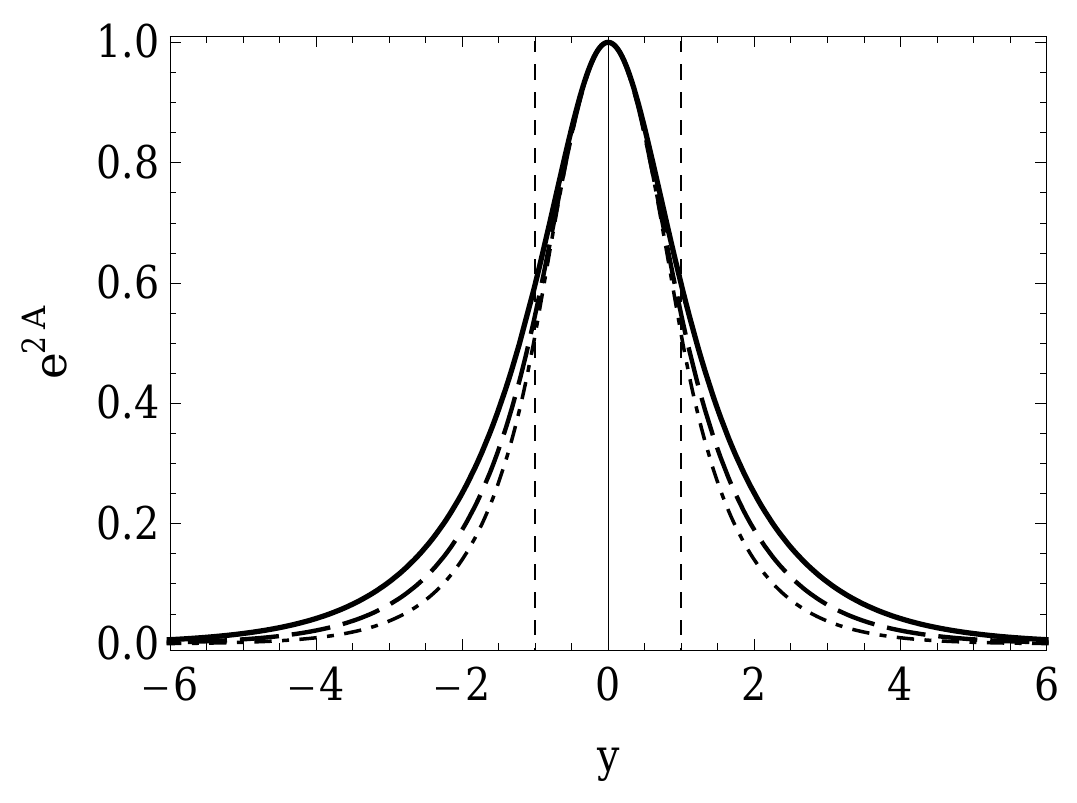}
    \end{center}
    \vspace{-0.9cm}
    \caption{\small{Warp factor $e^{2A(y)}$ plotted for same values of $n$ used in Fig.~\ref{fig6}.} \label{fig7}}
\end{figure}

The warp function is obtained using Eq. \eqref{WComp} in the second of Eqs. \eqref{FOF}. Although we can not invert Eq. \eqref{eqphicomp} to get the warp function in terms of $y$ analytically, given the fact that $\phi$ is always injective in $y$ and, consequently, an invertible function, we can still get the warp function in terms of the source field in the form
\begin{equation}\label{warpfacMB}
    A(\phi)= -\frac{\phi^2+2n\phi^2\times{}_2F_{1}\big(\frac1{n},\, 1,\, \frac{n+1}{n},\phi^{2n}\big)}{3(2n+1)}\,.
\end{equation}
In Fig. \ref{fig7} we show the behavior of warp factor, using the numerical solution of Eq. \eqref{eqphicomp}. In Ref. \cite{Bazeia:2014hja}, it was shows that the warp factor behaves differently inside and outside of the compact space, indicating that the brane has a hybrid structure. This can be seen by expanding Eq. \eqref{warpfacMB}, from which one obtains in the large $n$ limit
\begin{eqnarray}\nonumber
    A(y) = \left\{
    \begin{array}{clc}
        -y^2/3  &  \mbox{ to } &|y|\leq 1\,,\\
        -2|y|/3+1/3 & \mbox{ to }  &|y|>1\,.
    \end{array} \right.
\end{eqnarray}
Note that inside the compact space the behavior follows the expected for thick brane, whereas outside it corresponds to a thin brane scenario, which proves the hybrid structure of the brane.

\begin{figure}[!htb]
    \begin{center}
        \includegraphics[scale=0.6]{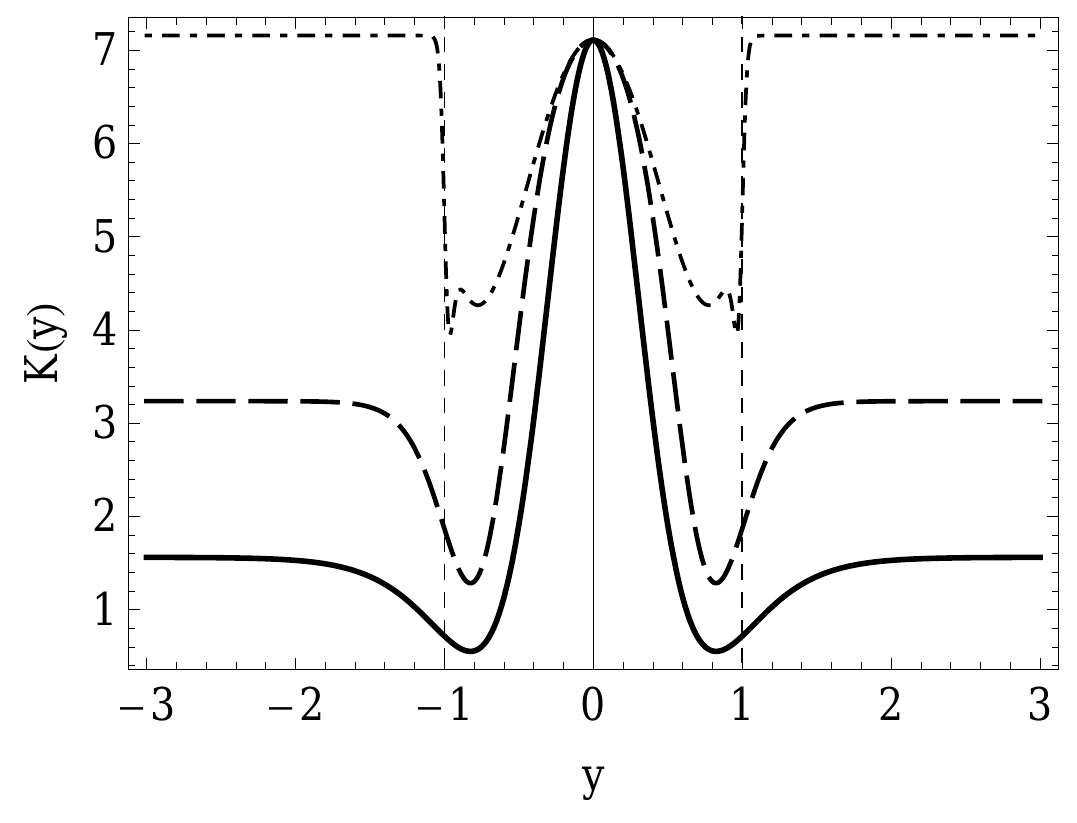}
    \end{center}
    \vspace{-0.9cm}
    \caption{\small{Kretschmann scalar plotted for $n$ as in Fig. \ref{fig6}.} \label{figure8}}
\end{figure}

Using Eqs.~\eqref{WComp} and \eqref{EqPOComp}, we can express the Kretschmann scalar in terms of the solution of the source field in the form,
\ben\nonumber
   K&=& \frac{64}{9} \big(1-\phi ^{2 n}\big)^4+\frac{640}{81} \Big(\phi -\frac{\phi ^{2 n+1}}{2 n+1}\Big)^4-\nn
   &&-\frac{256}{27} \big(1-\phi ^{2 n}\big)^2 \Big(\phi -\frac{\phi ^{2 n+1}}{2n+1}\Big)^2\,.
\een
In Fig. \ref{figure8} we show the behavior of Kretschmann scalar. Note that as $n$ increases, it becomes more and more confined in the compact space.

\begin{figure}[!htb]
    \begin{center}
        \includegraphics[scale=0.6]{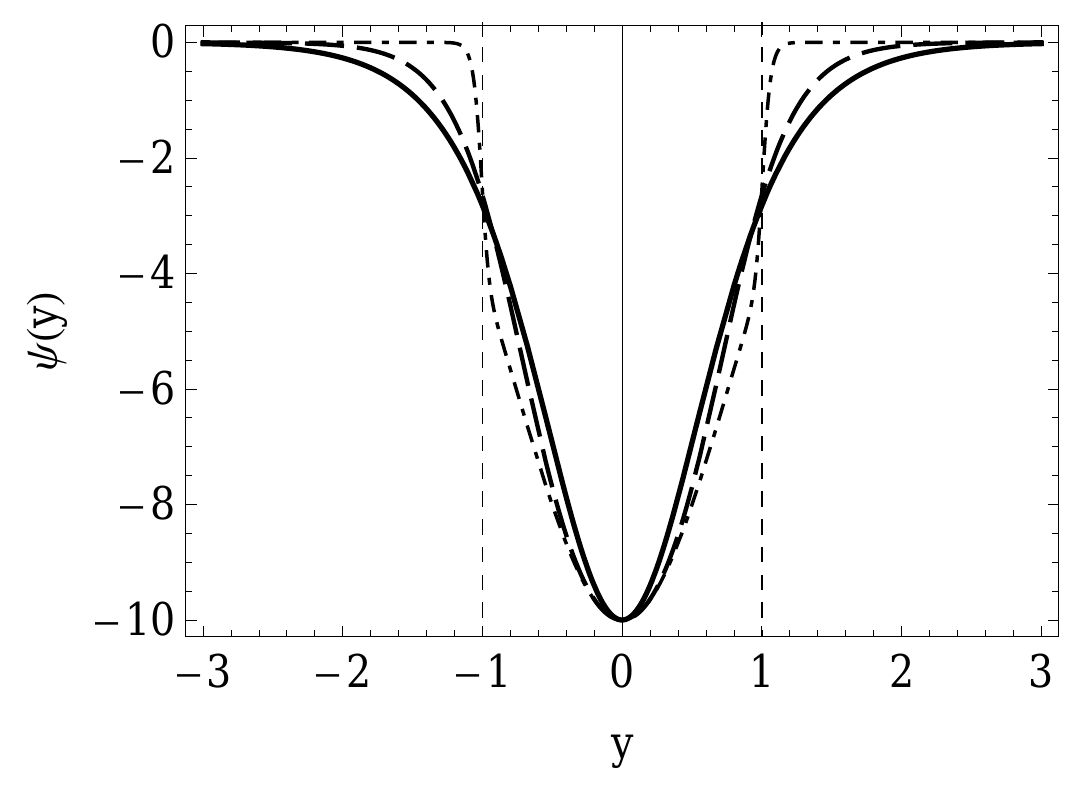}
    \end{center}
    \vspace{-0.9cm}
    \caption{\small{Plot of $\psi(y)$ giving by Eq. \eqref{PsiCompa} depicted for $\psi_0=-1$ and $n$ as in Fig. \ref{fig6}} \label{figure9}.}
\end{figure}

We now continue our analysis by extending the discussion for the auxiliary fields. First, let us use the first-order equations to rewrite the Eq. \eqref{eqpsi} as
\begin{eqnarray}
    9\psi'+2\left(8\phi+6 n \phi^{2n-1}-\frac{8 \phi^{2n+1}}{2n+1}\right)\psi=0.
\end{eqnarray}
As it was done for the warp factor, given the invertibility of $\phi$ as a function of $y$, we can use the chain-rule to write $\psi$ as a function of the source field, such that $\psi'=\psi_\phi \phi'$. With this, we can obtain the solution of $\psi$ analytically in terms of the source field, i.e.,
\begin{eqnarray}\label{PsiCompa}
\psi(\phi)=\psi_0\, \big(1-\phi^{2n}\big)^{2/3}e^{-8\,s(\phi)/9},
\end{eqnarray}
where
\begin{equation}\nonumber
   s(\phi)=\phi ^2 +\frac{2}{2n+1}\,B\big(\phi ^{2 n}; 1+\frac{1}{n},0\big),
\end{equation}
and $B\left(x; a,b\right)$ is the incomplete beta function defined as
\begin{equation}\nonumber
    B\big(x; a,b\big)=\int_{0}^xt^{a-1}(1-t)^{b-1}dt\,,
\end{equation}
where $a$ and $b$ are two parameters. In Fig. \ref{figure9} we show the profile of $\psi(y)$ for different values of $n$. We see that in the large $n$ limit, the field tends to become more and more confined in the compact region.

\begin{figure}[!htb]
    \begin{center}
        \includegraphics[scale=0.6]{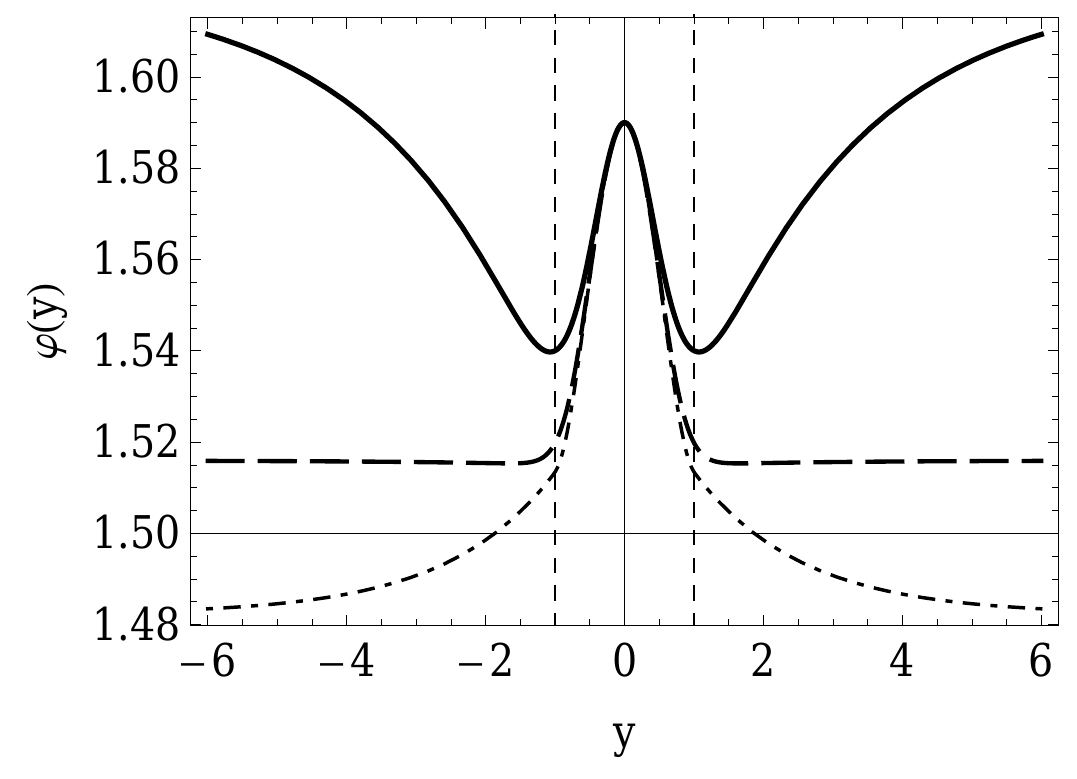}
    \end{center}
    \vspace{-0.9cm}
    \caption{\small{Numerical plot of $\varphi(y)$ for $\varphi'(0)=0$, $\psi_0=1$, and $\varphi_0=1.59$, with $n=1$ (solid line), $n=2$ (dashed line) and $n=20$ (dot-dashed line)}. \label{figure10}}
\end{figure}

\begin{figure}[!htb]
    \begin{center}
        \includegraphics[scale=0.6]{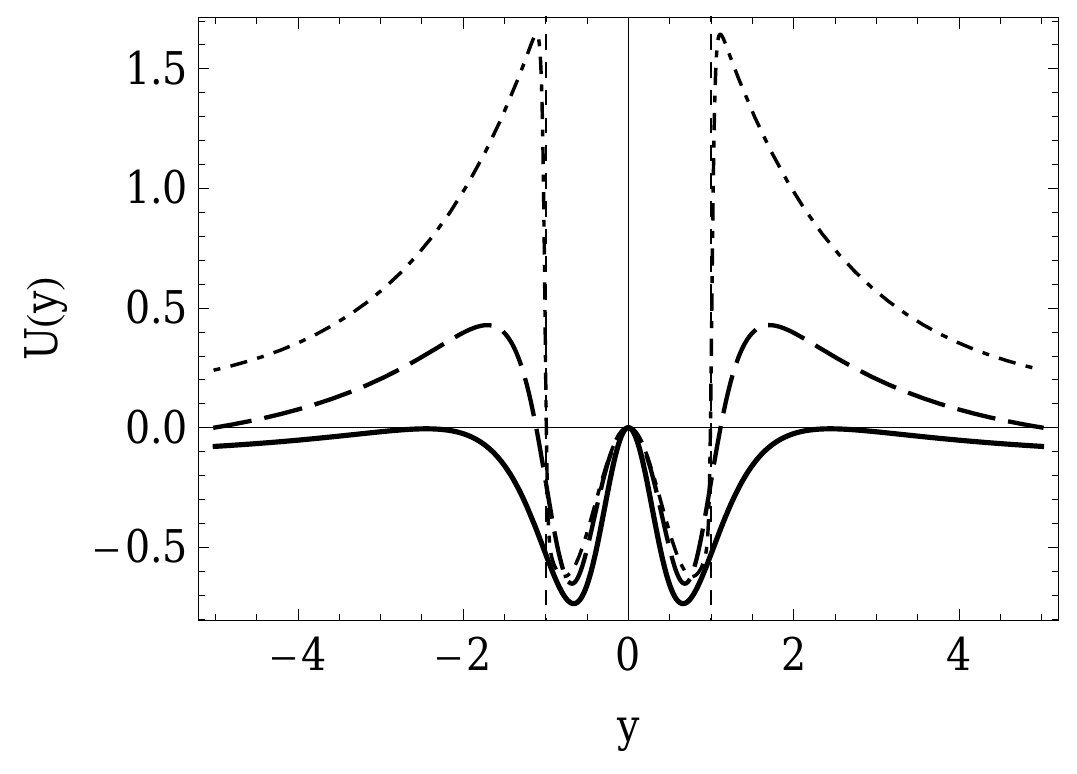}
    \end{center}
    \vspace{-0.9cm}
    \caption{\small{Numerical plot of $U(y)$ for $U(0)=0$, $\psi_0=1$, and $\varphi_0=1.54$, with $n=1$ (solid line), $n=2$ (dashed line) and $n=20$ (dot-dashed line)}. \label{figure11}}
\end{figure}

The auxiliary field $\varphi$ and the potential $U$ are again obtained via numerical methods subjected to the boundary conditions $\varphi(0)=\varphi_0$, $\varphi'(0)=0$, $U(0)=0$. Similarly to the previous case, depending on the values of $\psi_0$ and $\varphi_0$ the solution for the field $\varphi$ can feature a single minimum at $y=0$, a single peak at $y=0$, or a phase transition between these two values. However, there is a crucial difference between this model and the one presented in Sec.~\ref{modeloA}. For the previous model, for a fixed value of $\psi_0$, this phase transition between the two behaviors occurred  for a fixed range of $\varphi_0$, independently of the free parameter $a$. However, in this model the range of values of $\varphi_0$ for which the phase transition occurs is also controlled by the parameter $n$, which means that even for a specific combination of $\varphi_0$ and $\psi_0$ one can have different qualitative behaviors of the field $\varphi_0$ by varying the parameter $n$. These results are plotted in Fig. \ref{figure10}. Regarding the potential $U$, the situation is similar. The possible qualitative behaviors of $U$ are the same as in the previous model in Sec.~\ref{modeloA}, but for a fixed $\psi_0$ the range of values for $\varphi_0$ for which the phase transition occurs is strongly dependent on $n$, being thus possible to chose a specific combination of $\psi_0$ and $\varphi_0$ and still have different qualitative behaviors of $U$. These results are shown in Fig.~\ref{figure11}.

 As a next model, we will propose a different situation, where the compactification process will be more concentrated on the geometry rather than the source field.

\subsection{Third model}\label{modeloC}

Let us now examine a third model that allows for important modifications in the geometry of the brane and how this interferes with the auxiliary fields. To do this, we consider a $W(\phi)$ of the form,
\begin{equation}\label{superMC}
    W(\phi)=-\frac1{\sqrt{\lambda}(1-\lambda)}\ln\Big[\frac{1-\sqrt{\lambda}\,\sn(\phi,\lambda)}{\dn(\phi,\lambda)}\Big],
\end{equation}
where $\lambda$ is a real parameter which is in the range $[0,1)$. The functions $\sn(\phi,\lambda)$ and $\dn(\phi,\lambda)$ are the Jacobi's elliptic functions. For $\lambda=0$ we have the usual trigonometric functions, and for $\lambda= 1$ we recover the hyperbolic functions. In Ref. \cite{Bazeia:2015eta} the authors showed that the parameter $\lambda$ significantly affects the compactification of the warp factor.

\begin{figure}[!htb]
    \begin{center}
        \includegraphics[scale=0.6]{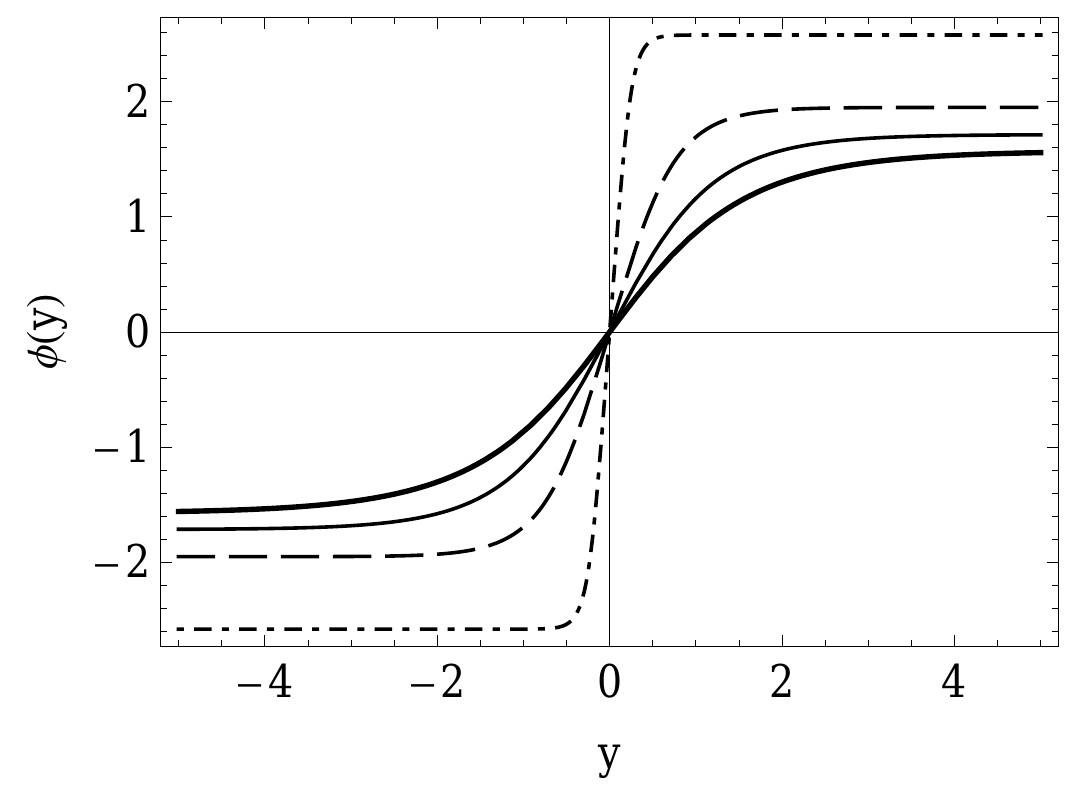}
    \end{center}
    \vspace{-0.9cm}
    \caption{\small{Kink solution given by Eq. \eqref{eqsolModC}, plotted for $\lambda=0$ (thick-solid line), $\lambda=0.3$ (thin-solid line), $\lambda=0.6$ (dashed line) and $\lambda=0.9$ (dot-dashed line).} \label{figure12}}
\end{figure}

Using the first-order formalism, we find the solution for the source field $\phi$ from the first of Eq. \eqref{FOF} as
\begin{equation}\label{eqsolModC}
    \phi(y)=\sn^{-1}\left[\tanh\bigg(\frac{y}{1-\lambda}\bigg),\,\lambda\,\right].
\end{equation}
It can be shown that we obtain the model analyzed in \ref{modeloA} when $\lambda=0$. On the other hand, if $\lambda\neq 0$ we have a qualitative difference with respect to the previously studied model. In this case, the asymptotic value of the solution change for $\phi(y\to\pm\infty)\to\pm \phi_v$, where,
\begin{equation}\nonumber
    \phi_v=\,K(\lambda)\,,
\end{equation}
where $K\left(\lambda\right)$ is the complete elliptic integral of the first kind (not to be confused with the Kretschmann scalar). For $\lambda=0$ we have $K(0)=\pi/2$ and, on the other hand, if $\lambda\to 1$ we have $K(\lambda\to 1) \to \infty$. Fig. \ref{figure12} shows the solution of field given  by Eq.  \eqref{eqsolModC} for $\lambda=0,\, 0.3,\, 0.6,\, 0.9$. One notices that $\lambda$ changes the thickness of the topological defect so that it becomes more and more concentrated at the origin as $\lambda$ increases towards unity. However, in this case we also have a shift in the asymptotic values of the field configuration.

We can also use the first-order formalism to express the scalar potential $V$ given by the Eq. \eqref{PotV} as
\ben
    V(\phi)&=&-\frac{\cn^2(\phi ,\lambda ) \Big[\dn^2(\phi ,\lambda )+2 \sqrt{\lambda} \,\sn(\phi ,\lambda )-2\Big]}{2 (\lambda-1)^2 \, \dn^2(\phi ,\lambda )\big(\sqrt{\lambda } \,\sn(\phi ,\lambda )-1\big)^2}-\nn
    &&-\frac{4}{3\lambda(\lambda-1)^2}\Big[\ln\Big(\frac{\dn(\phi ,\lambda )}{1-\sqrt{\lambda} \,\sn(\phi ,\lambda )}\Big)\Big]^2.
\een
Using the asymptotic value of the source field solution one verifies that the asymptotic value of the potential is given by
\begin{equation}\nonumber
    V(\phi\to\pm\phi_v)=-\frac{1}{3\lambda(\lambda-1)^2}\Big[\ln\frac{\!\!1-\lambda}{\,\,\big(1-\sqrt{\lambda}\big)^2}\Big]^2.
\end{equation}
Note that since $\lambda$ is in the range $[0,1)$, the asymptotic value of the potential is negative. We know that this result leads to an $AdS_5$ brane, since the cosmological constant is defined as $\Lambda_5\equiv V(\phi\to\phi_v)<0$ in five dimensions. When the parameter $\lambda$ tends to zero we have that $\Lambda_5\to-4/3$, and when $\lambda$ tends to one we have $\Lambda_5\to-\infty$.

\begin{figure}[!htb]
    \begin{center}
        \includegraphics[scale=0.6]{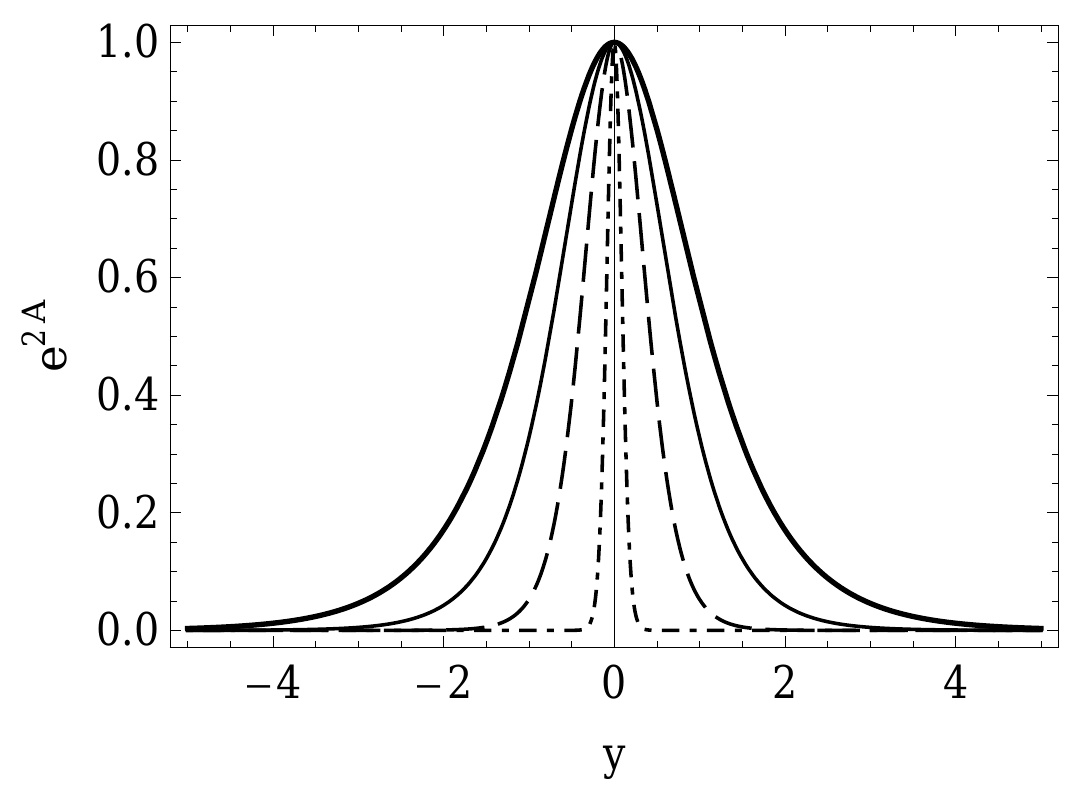}
    \end{center}
    \vspace{-0.9cm}
    \caption{\small{Warp factor plotted for $\lambda=0$ (thick-solid line), $\lambda=0.3$ (thin-solid line), $\lambda=0.6$ (dashed line) and $\lambda=0.9$ (dot-dashed line).} \label{figure13}}
\end{figure}

The warp factor is obtained numerically from the second of Eq. \eqref{FOF} is shown in Fig. \ref{figure13}, where we have used the same values for $\lambda$, i.e., $\lambda=0,\, 0.3,\, 0.6,\, 0.9$. Note that unlike the previous model \ref{modeloB}, the warp factor undergoes a strong compression with the increase of the parameter $\lambda$, with its thickness rapidly approaching zero as $\lambda$ gets closer and closer to one. This result suggests that this compactification in the geometry is much more intense than that described in the previous sections.

We also verify that the Kretschmann scalar presents a standard behavior. In this case, $K$ is given in terms of $\phi$ as
\ben
K(\phi)&=&\frac{512\cn^2(\phi,\lambda)\big[2-\dn^2(\phi,\lambda)-2\sqrt{\lambda }\, \sn(\phi,\lambda) \big]}{81\lambda (\lambda-1)^4 \dn^2(\phi,\lambda) \big[\sqrt{\lambda}\,\sn(\phi,\lambda)-1\big]^2}\times\nn
&&\times\left[\ln\Big(\frac{\dn(\phi,\lambda)}{1-\sqrt{\lambda}\sn(\phi,\lambda)\,}\Big)\right]^2+\nn
&&+\frac{5}{81\lambda^2\, (\lambda -1)^4 }\left[\ln\Big(\frac{\dn(\phi,\lambda)}{1-\sqrt{\lambda}\,\sn(\phi,\lambda)\,}\Big)\right]^4+\nn
&&+\left[\frac{4\, \cn(\phi,\lambda)}{3\,(\lambda -1) \dn(\phi,\lambda) }\right]^4.
\een
The asymptotic limits of the above expression are
\begin{equation}\nonumber
   \lim_{y\to 0}K(y)=\frac{64}{9(\lambda-1)^4}\,,
\end{equation}
and
\begin{equation}\nonumber
    \lim_{y\to \pm\infty}K(y)=\frac{40}{81\lambda^2(\lambda-1)^4}\Big[\ln\frac{\!\!1-\lambda}{\,\,\big(1-\sqrt{\lambda}\big)^2}\Big]^4\,.
\end{equation}
Note that when $\lambda\to 0$ we have $K(0)\to64/9$ and $K(y\to\pm\infty)\to640/81$. On the other hand, if $\lambda\to 1$ we have $K(0)=K(y\to\pm\infty)\to\infty$, suggesting that the limit $\lambda=1$ is unreachable, which was already taken into account when this value was excluded from the interval range of $\lambda$.

Let us now analyze how the model proposed in this section modifies the auxiliary fields. Using the function given by Eq. \eqref{superMC} into Eq. \eqref{eqpsi}, one obtains a differential equation that can be solved numerically for $\psi$,
\ben\nonumber
 9\psi'\!+\!6\frac{\sn(\phi,\lambda)}{\dn^2(\phi,\lambda)}\psi\!+\!\frac{16\psi}{\sqrt{\lambda}(1\!-\!\lambda)}\ln\!\Big[\frac{\dn(\phi,\lambda)}{1\!-\!\sqrt{\lambda}\,\sn(\phi,\lambda)}\Big]\!\!=\!0.\nonumber
\een

\begin{figure}[!htb]
    \begin{center}
        \includegraphics[scale=0.6]{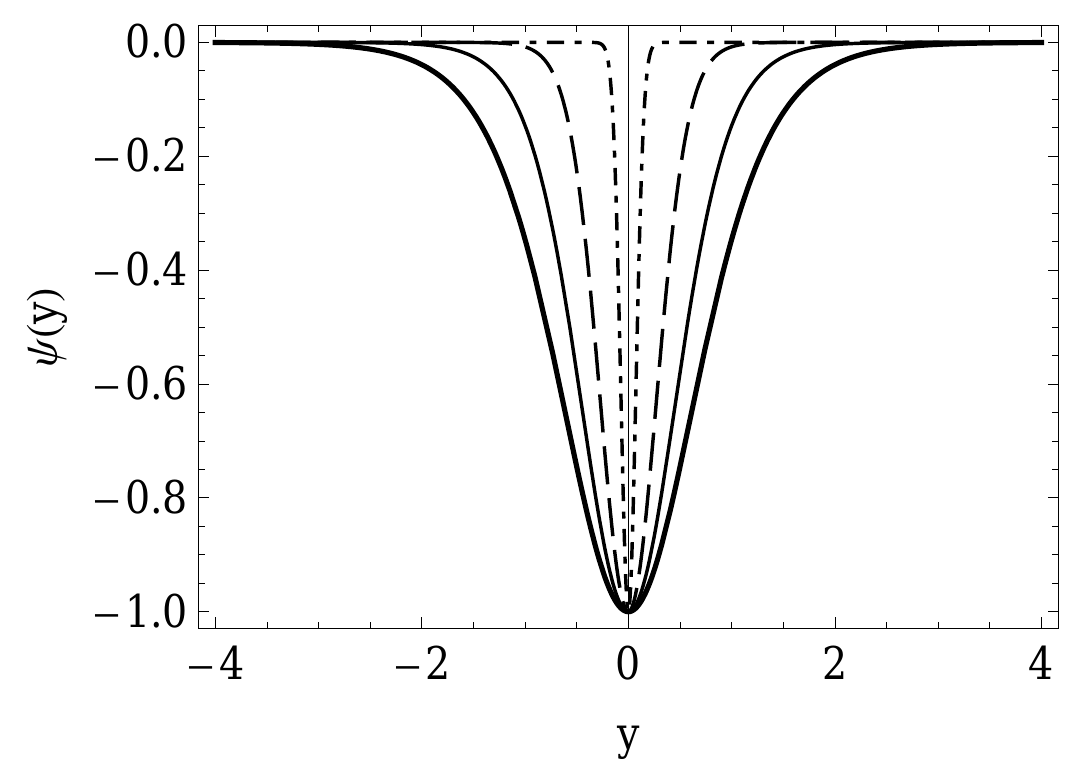}
    \end{center}
    \vspace{-0.9cm}
    \caption{\small{Plot of $\psi(y)$ for $\psi_0=-1$, $\lambda=0$ (thick-solid line), $\lambda=0.3$ (thin-solid line), $\lambda=0.6$ (dashed line) and $\lambda=0.9$ (dot-dashed line).} \label{figure14}}
\end{figure}

The result of this calculation is shown in Fig. \ref{figure14}. It is visible that the compactification is also quite accentuated with an increase in $\lambda$, corresponding to a consequent decrease in the thickness of the field.

\begin{figure}[!htb]
    \begin{center}
        \includegraphics[scale=0.6]{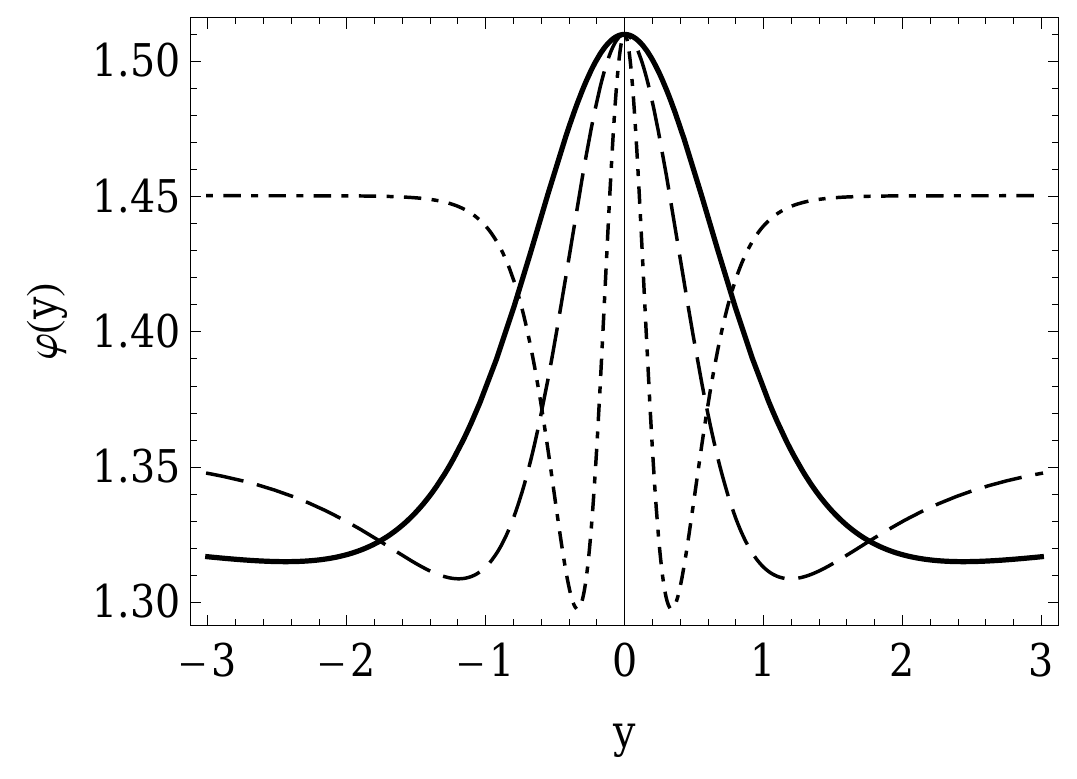}
    \end{center}
    \vspace{-0.9cm}
    \caption{\small{Numerical plot of $\varphi(y)$ for $\psi_0=1$ and $\varphi_0=1.51$, with $\lambda=0$ (solid line), $\lambda=0.4$ (dashed line), and $\lambda=0.8$ (dot-dashed line)}. \label{figure15}}
\end{figure}

\begin{figure}[!htb]
    \begin{center}
        \includegraphics[scale=0.6]{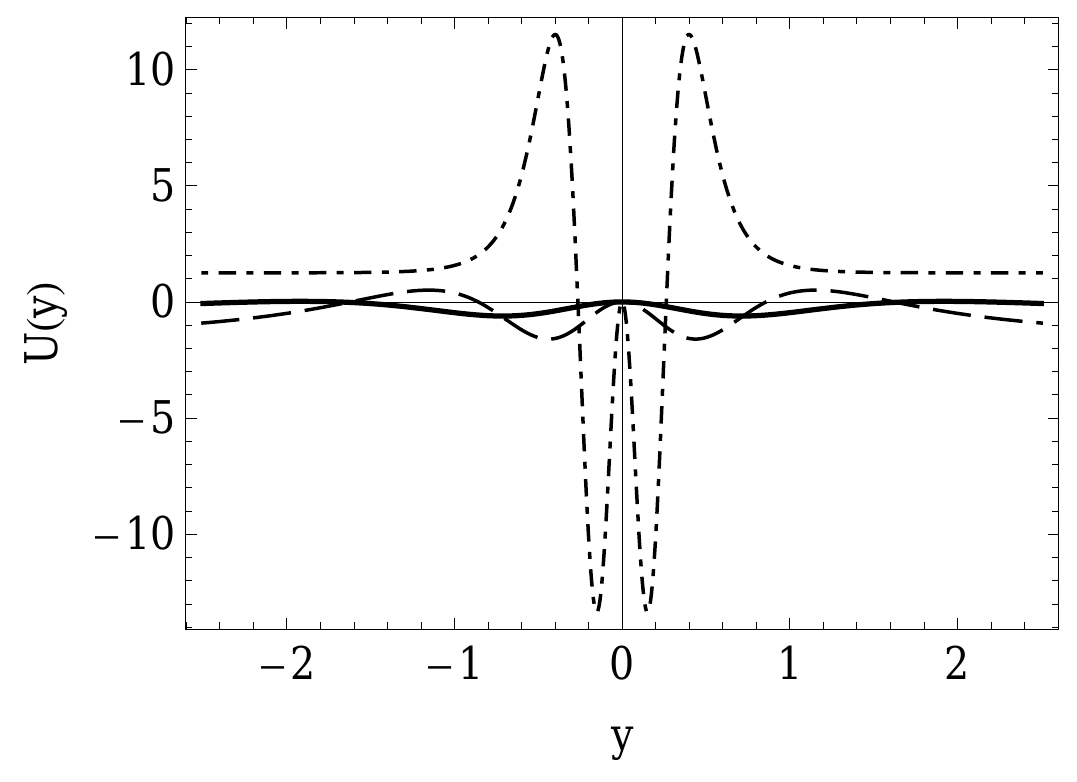}
    \end{center}
    \vspace{-0.9cm}
    \caption{\small{Numerical plot of $U(y)$ for $U(0)=0$, $\psi_0=1$, and $\varphi_0=1.47$, with $\lambda=0$ (solid line), $\lambda=0.4$ (dashed line), and $\lambda=0.8$ (dot-dashed line)}. \label{figure16}}
\end{figure}

The field $\varphi$ and the potential $U$ must again be obtained via numerical methods, subjected to the boundary conditions $\varphi(0)=\varphi_0$, $\varphi'(0)=0$, and $U(0)=0$. Similarly to the previous cases studied, the general behavior of the field $\varphi$ can be either a global minimum and outwards increase, a global maximum and an outwards decrease, or some transition state between these two behaviors. In this model, the compactification parameter $\lambda$ also influences the qualitative behavior of this field, which means that even for a specific combination of $\varphi_0$ and $\psi_0$ the qualitative behavior of $\varphi$ might change depending on the value of $\lambda$, and thus this parameter has a stronger influence on the solutions than the compactification parameter $a$ in Sec. \ref{modeloA}. In Fig. \ref{figure15}, we plot solutions for $\varphi$ with different values of $\lambda$. As for the solutions for the potential $U$, similarly as the previous models, the behaviors of this function can again be either a local minimum at $y=0$ surrounded by two maxima and an outwards decrease, a local maximum surrounded by two minima and an outwards increase, or some transition phase between the two, depending on the combination of parameters chosen. For a fixed combination of $\psi_0$ and $\varphi_0$, the parameter $\lambda$ also influences strongly the shape of the potential $U$, allowing for different qualitative behaviors. These results are plotted in Fig.~\ref{figure16}.


\subsection{Fourth model}\label{modeloD}

Let us now focus on a different issue concerning the study of braneworld scenarios in the presence of asymmetry. This possibility was explored before in \cite{AS0,AS1,AS2,AS3,AS4,AS5,AS6,AS7,AS8} in several distinct situations. Here, we want to examine how the asymmetry works in the scalar-tensor representation of the $f(R,T)$ brane, and how it modifies the behavior of the fields $\varphi$ and $\psi$. Since this is a first approach to this issue, we shall work on a simpler setting without taking into consideration the effects of compactification of the source field $\phi$. To implement this, let us take the function $W(\phi)$ to be
\begin{equation}\label{WmodD}
    W(\phi)=c+\phi-\frac{1}{3}\phi^{3},
\end{equation}
where the real constant $c$ must belong in the range $-2/3<c<2/3$ to preserve the physical relevance of the solutions. This model reduces to the one studied in Sec. \ref{modeloB} with $n=1$ in the particular case of $c=0$, and here we add $c$ to build the asymmetric scenario; see, e.g., Ref. \cite{AS4}.

\begin{figure}[!htb]
    \begin{center}
        \includegraphics[scale=0.6]{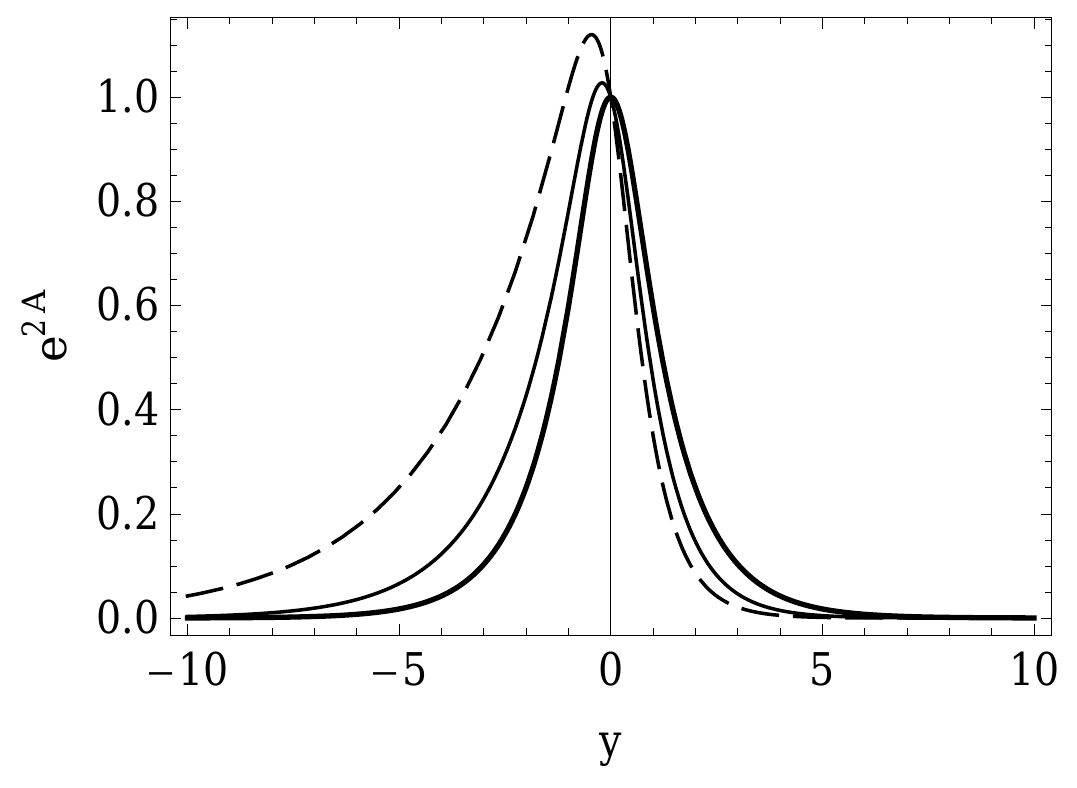}
    \end{center}
    \vspace{-0.9cm}
    \caption{\small{Warp factor $e^{2A}$ given by Eq. \eqref{warpMD} for $c=0$ (thick solid line), $c=0.2$ (thin solid line) and $c=0.4$ (dashed line).} \label{figure17}}
\end{figure}

We can verify that $c$ does not modify the source field solution, which in this case can be obtained by the first-order formalism as $\phi(y)=\tanh(y)$. On the other hand, the parameter $c$ changes in the warp function, which can be written as
\begin{equation}\label{warpMD}
    A(y)=-\frac{2c}{3}y+\frac{4}{9} \ln[\sech(y)]-\frac19\tanh^2(y).
\end{equation}
In Fig. \ref{figure17} we show the warp factor obtained in Eq. \eqref{warpMD} for $c=0,\,0.2,\,0.4$. We verify that in the parameter region considered for $c$ the warp factor becomes asymmetric but remains localized, vanishing asymptotically. The Kretschmann scalar is also influenced by this parameter, such that
\ben
K&=&\!\frac{64}{81} \left(9-12 c^2+10 c^4\right)+\frac{512}{81} c \left(5 c^2-3\right) \tanh(y)+\nn
&&\!+\frac{256}{27}\! \left(7 c^2\!-\!4\right)\! \tanh^2(y)\!+\!\frac{512}{243} c\! \left(36\!-\!5 c^2\right)\! \tanh^3(y)\!+\!\nn
&&\!+\frac{256}{81} \left(24-13 c^2\right) \tanh^4(y)-\frac{5120}{81} c \tanh^5(y)+\nn
&&\!+\frac{256}{243} \left(5 c^2-59\right) \tanh^6(y)+\frac{4096}{243} c \tanh^7(y)+\nn
&&\!+\frac{5056 }{243}\tanh^8(y)\!-\!\frac{2560 c }{2187}\tanh^9(y)-\nn
&&-\frac{4864 }{2187}\tanh^{10}(y)\!+\frac{640 }{6561}\tanh^{12}(y).
\een
So, $K(0)=64(9-12c^2+10c^4)/81$ and $K(y\to\pm\infty)=640(2\pm 3c)^4/6561$. Also, for $c\neq0$ the brane connects two different $AdS_5$ spaces, with the cosmological constants becoming $\Lambda_5\equiv V(\phi\to\pm\phi_v)=-4(2\pm3c)^2/27$.

\begin{figure}[!htb]
    \begin{center}
        \includegraphics[scale=0.6]{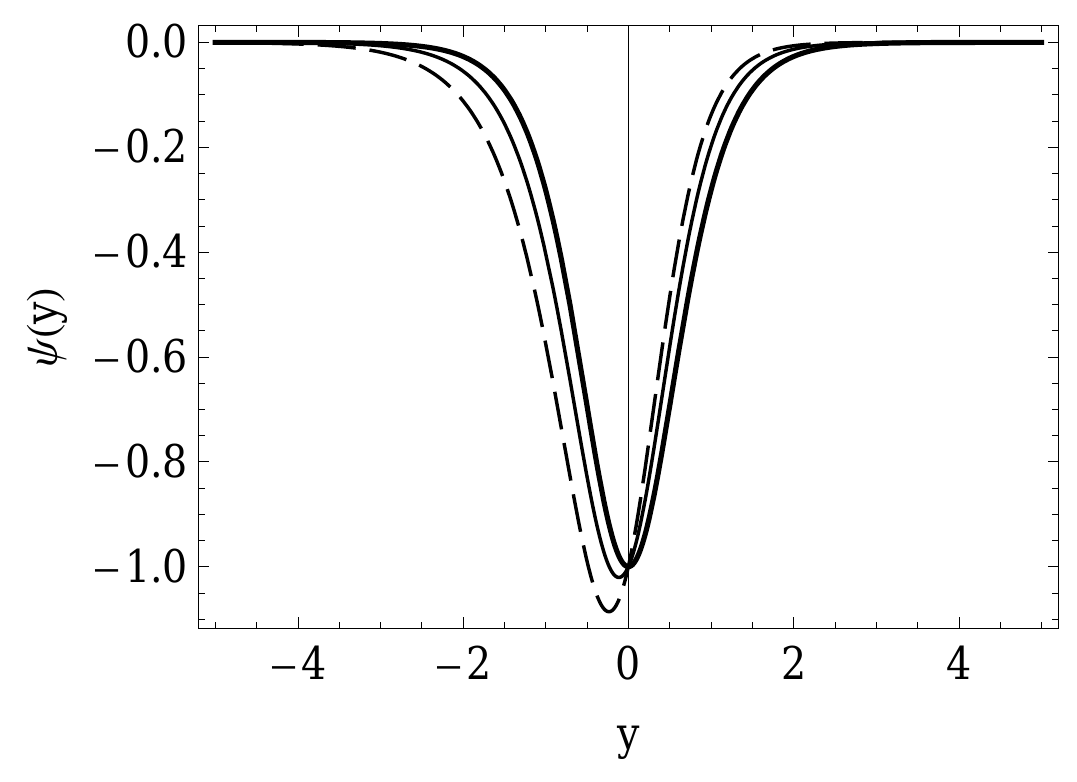}
    \end{center}
    \vspace{-0.9cm}
    \caption{\small{The field $\psi(y)$ given by Eq.\eqref{AuxiPsiMD} depicted for $\psi_0=-1$ and $c=0$ (thick solid line), $c=0.2$ (thin solid line) and $c=0.4$ (dashed line).} \label{figure18}}
\end{figure}

Taking Eq. \eqref{WmodD} into Eq. \eqref{eqpsi} and taking the source field solution $\phi(y)=\tanh(y)$ one obtains a differential equation for $\psi$ as
\begin{equation}\nonumber
    9\psi'(y)+\frac43\Big(12 c + 21 \tanh(y) - 4 \tanh^3(y)\Big)\psi(y)=0.
\end{equation}
This equation can be solved analytically and yields a solution of the form
\begin{equation}\label{AuxiPsiMD}
    \psi(y)=\psi_0 \frac{e^{-8\tanh^2(y)/27}}{\cosh^{68/27}(y)}e^{-16cy/9}.
\end{equation}
Fig. \ref{figure18} shows this auxiliary field $\psi$ for $\psi_0=-1$ and $c=0,\,0.2,\,0.4$. Note that for $c\neq 0$ the minima value of $\psi(y)$ is displaced from $y=0$, but the asymptotic values remain the same. This exact result show that the parameter $c$ induces an asymmetric behavior in the auxiliary field $\psi$. We verified that the results are qualitatively similar for negative values of $c$.

\begin{figure}[!htb]
    \begin{center}
        \includegraphics[scale=0.6]{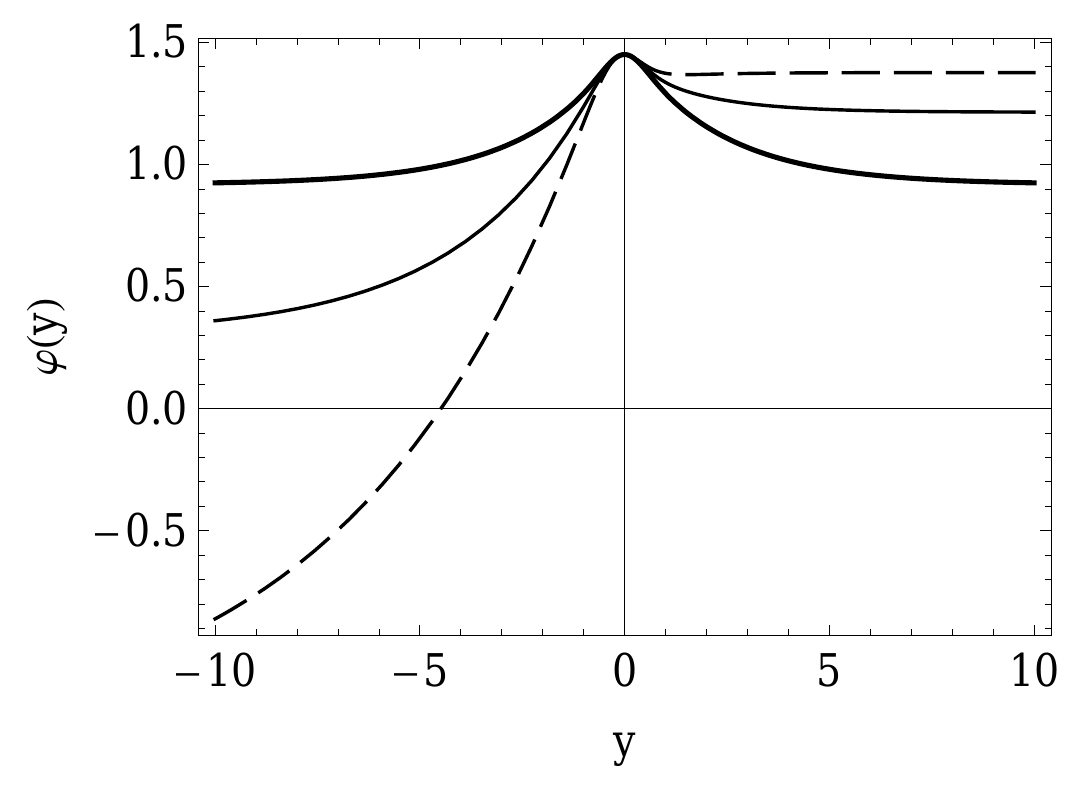}
    \end{center}
    \vspace{-0.9cm}
    \caption{\small{Plot of $\varphi(y)$ for $\psi_0=1$ and $\varphi_0=1.45$, with $c=0$ (thick solid line), $c=1$ (thin solid line), and $c=0.4$ (dashed line)}. \label{figure19}}
\end{figure}

\begin{figure}[!htb]
    \begin{center}
        \includegraphics[scale=0.6]{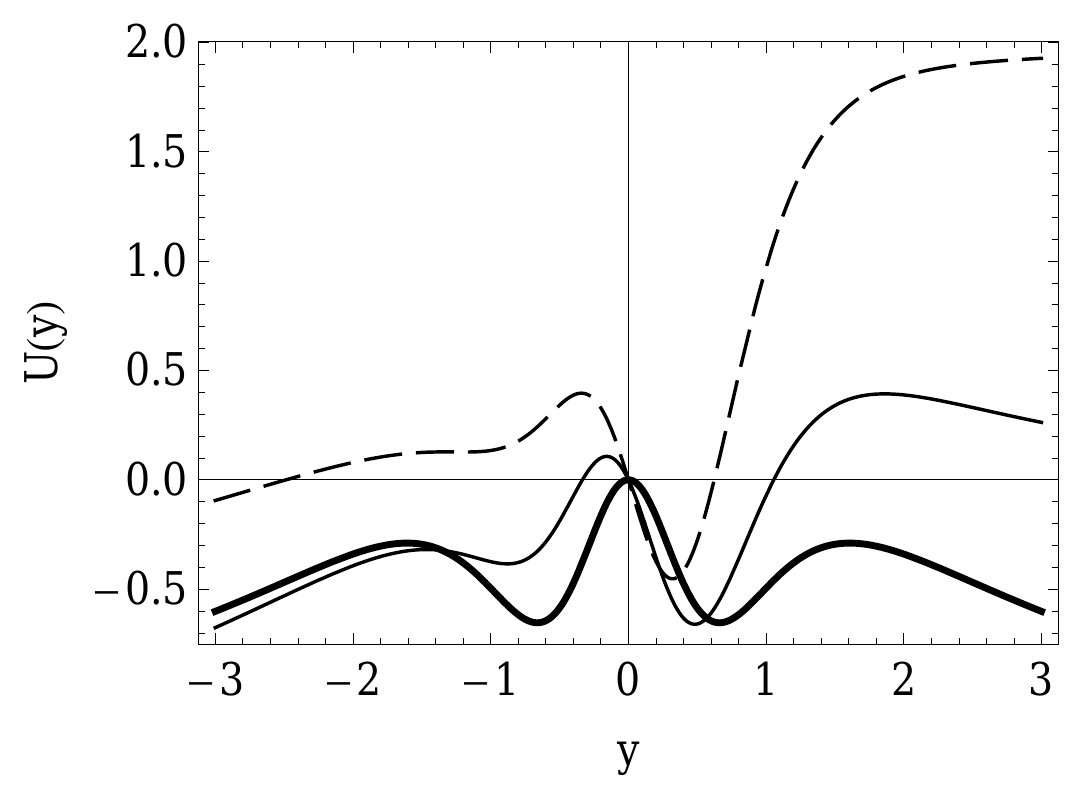}
    \end{center}
    \vspace{-0.9cm}
    \caption{\small{Plot of $U(y)$ for $U(0)=0$, $\psi_0=1$, and $\varphi_0=1.45$, with $c=0$ (thick solid line), $c=1$ (thin solid line), and $c=0.4$ (dashed line)}. \label{figure20}}
\end{figure}

Let us now verify how the auxiliary field $\varphi$ responds to variations in $c$. To do so, we take Eqs. \eqref{eqvarphi} and \eqref{eqU} with the function \eqref{WmodD} and solve them numerically. For $c=0$, the general behaviors of the scalar field $\varphi$ are the same as in the previous models, i.e., either a single peak at $y=0$, a single minimum at $y=0$, or some transition phase between the two. However, as $c$ increases, the induced asymmetry can radically change the behavior of $\varphi$. For a single-peak solution with $c=0$, an increase in $c$ can raise one of the asymptotic values above the central value, which ceases to be a global maximum. For $c$ large enough, this local maximum can even disappear and the field $\varphi$ becomes monotonically increasing. A similar analysis holds for the single minimum behavior. These results are plotted in Fig. \ref{figure19}. Regarding the potential $U$, for $c=0$ the general possible behaviors are the standard, i.e., either a local minimum at $y=0$ surrounded by two maxima and an outwards decrease, a local maximum surrounded by two minima and an outwards increase, or a transition state between the two. However, again the asymmetry induced by the parameter $c$ can produce clear changes in the potential $U$, leading to complicated behaviors with sequences of peaks and droughts of different heights. These results are plotted in Fig. \ref{figure20}. In both cases we use $c$ as in Fig. \ref{figure17} and also the set of initial conditions: $\psi_0=1$ and $\varphi_0=1.45$. Note that the asymptotic behaviors also depends on the parameter $c$.

\section{Tensor perturbations}\label{secStability}

In this section, we will perform a linear stability analysis of the gravitational sector considering linear perturbations in the metric $g_{ab}$ and in the scalar field $\phi$ in the form $\eta_{\mu\nu}\to \eta_{\mu\nu}+h_{\mu\nu}(r,y)$ and $\phi\to\phi(y)+\xi(r,y)$, where $r$ represents the four-dimensional position vector. The perturbed metric is given by
\begin{equation}
    g_{ab}=e^{2A}\big(\eta_{\mu\nu}+h_{\mu\nu}\big)dx^{\mu}dx^{\nu}-dy^2,
\end{equation}
where $h_{\mu\nu}$ satisfies the transverse and traceless (TT) conditions $\partial^{\mu} h_{\mu\nu}=0$ and $\eta^{\mu\nu}h_{\mu\nu}=0$. Furthermore, it was show in Ref. \cite{Bazeia:2015owa} that if the function $f(R,T)$ in Eq. \eqref{actiongeo} is separable in the form $f_1(R)+f_2(T)$, the perturbative equation of the geometric sector decouples from the perturbation of the field $\phi$. In the scalar-tensor representation this condition is identical to considering the potential $U(\psi,\varphi)=U_1(\psi)+U_2(\varphi)$. With this prescription, the equation for the perturbation $h_{\mu\nu}$ obtained from Eq. \eqref{fieldst} can be written as
\begin{equation}\nonumber
\Big(\!-\!\partial_y^2\! -\!4A^\prime \partial_y\!+\!e^{-2A}\Box^{(4)} \!-\!\frac{\varphi^\prime}{\varphi}\partial_y\!\Big)h_{\mu\nu}=0 \,.
\end{equation}
We can write $h_{\mu\nu}$ in terms of a new $z$-coordinate defined in terms of the $y$-coordinate as $dz=e^{-A(y)}dy$, and also perform a redefinition of the perturbation in the metric of the form $H_{\mu\nu}= e^{ipr}e^{3A(z)/2}\varphi^{1/2}h_{\mu\nu}$. Under these redefinition, the stability of the gravitational sector is determined by a Schrodinger-like equation in the form
\ben\label{eq49}
\Big[-\frac{d^2}{dz^2}+{\cal U}(z)\Big]H_{\mu\nu}= p^2H_{\mu\nu}\,,
\een
where the potential ${\cal U}$ is written as
\ben\label{eq50}
{\cal U}(z)=\alpha^2(z) -\frac{d\alpha}{dz}\,,
\een
and the function $\alpha(z)$ is defined as
\begin{equation}\nonumber
    \alpha(z)=-\frac32\frac{dA}{dz}-\frac12\frac{d}{dz}\big(\ln\varphi\big)\,.
\end{equation}

Equation \eqref{eq49} can be factorized in the form $S^{\cal y} S\,H_{\mu\nu}= p^2H_{\mu\nu}$, where $ S^{\cal y}=-d/dz+\alpha(z)$ and $p^2\geq 0$. Therefore, given the positivity of the eigenvalue $p^2$, the theory remains stable against tensor perturbations. The massless graviton state represented by the zero-mode is then
\ben\label{eq51}
{H}_{\mu\nu}^{(0)}(z)=N_{\mu\nu} \sqrt{\varphi(z)}\,e^{3A(z)/2}\,,
\een
where $N_{\mu\nu}$ is a normalization factor. Following construction conducted in this section, one verifies that the field $\varphi$ must always be positive to guarantee that the perturbations in the metric remain real, see Eq. \eqref{eq51}. This condition must constrain the values of the initial conditions that we can use to solve the differential equations Eqs. \eqref{eqpsi} to \eqref{eqU}. Moreover, a transformation back to the variable $y$ leads to a form of the stability potential as
\ben
{\cal U}(y)&=&e^{2A}\Big[2A'\big(\!\ln\varphi\big)^{\prime}\!+\frac14\big(\!\ln\varphi\big)^{\prime 2}\!+\frac12 \big(\!\ln\varphi\big)^{\prime\prime}\Big]+\nn
&&+\frac{3}4e^{2A}\big(5 A'^2+2 A''\big).\label{eqPotEsy}
\een

In what follows, we analyze the behavior of the stability potential given by Eq. \eqref{eqPotEsy} and the graviton zero-mode in Eq. \eqref{eq51}, for the four models studied in this work.

\subsection{Stability of the first model}

\begin{figure}[!htb]
    \begin{center}
        \includegraphics[scale=0.37]{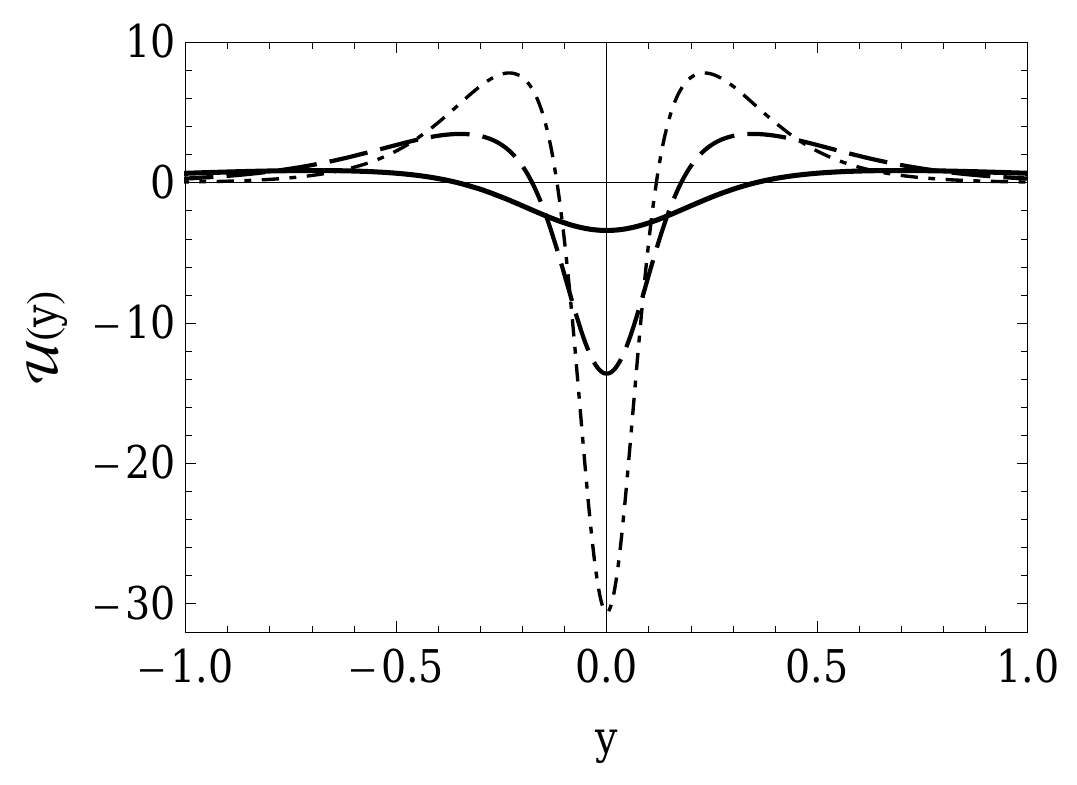}
        \hspace{0.2 cm}
        \includegraphics[scale=0.37]{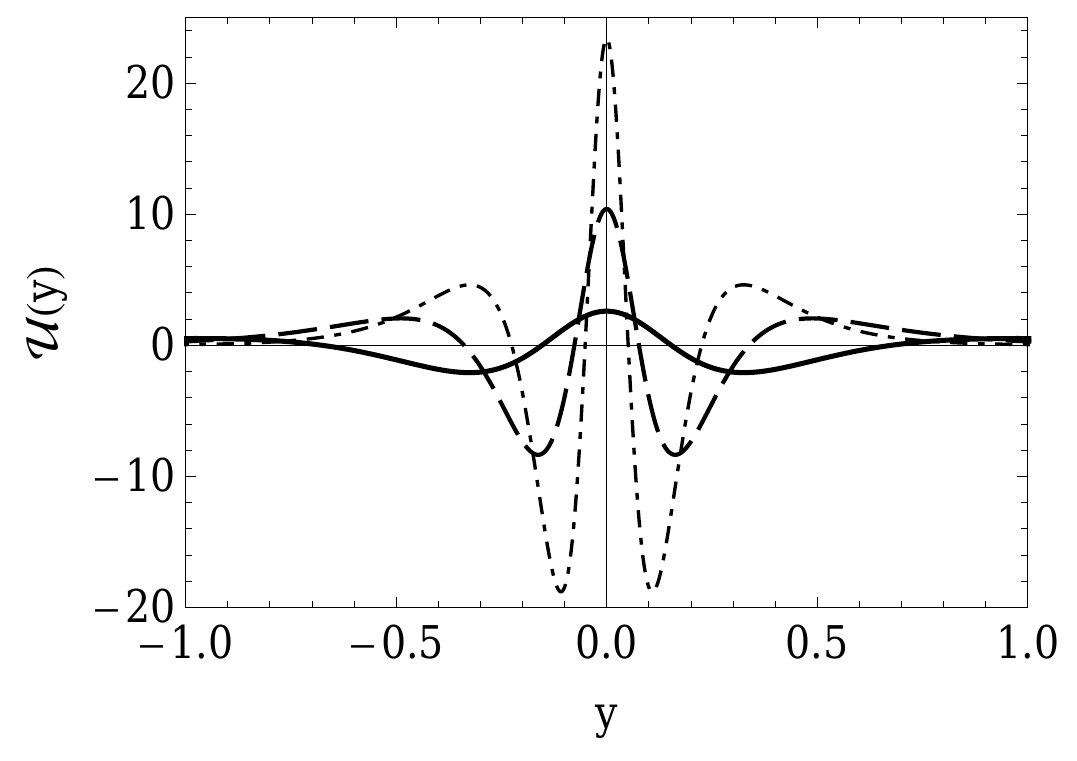}
        \includegraphics[scale=0.37]{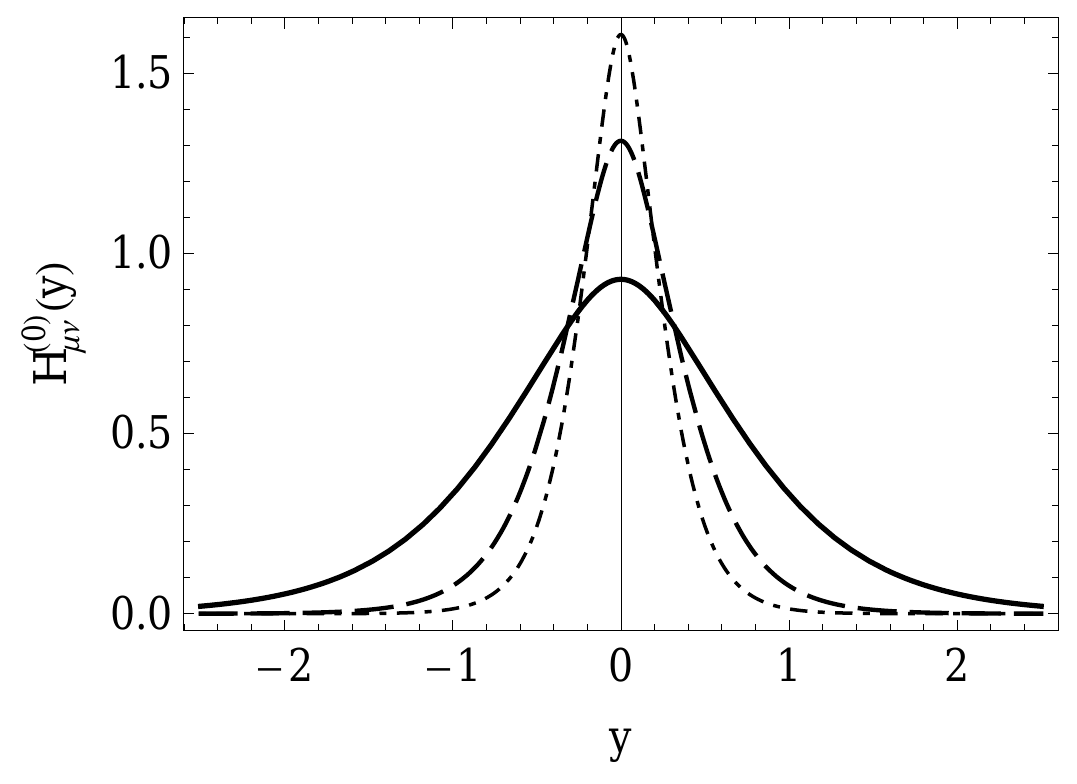}
        \hspace{0.2 cm}
        \includegraphics[scale=0.37]{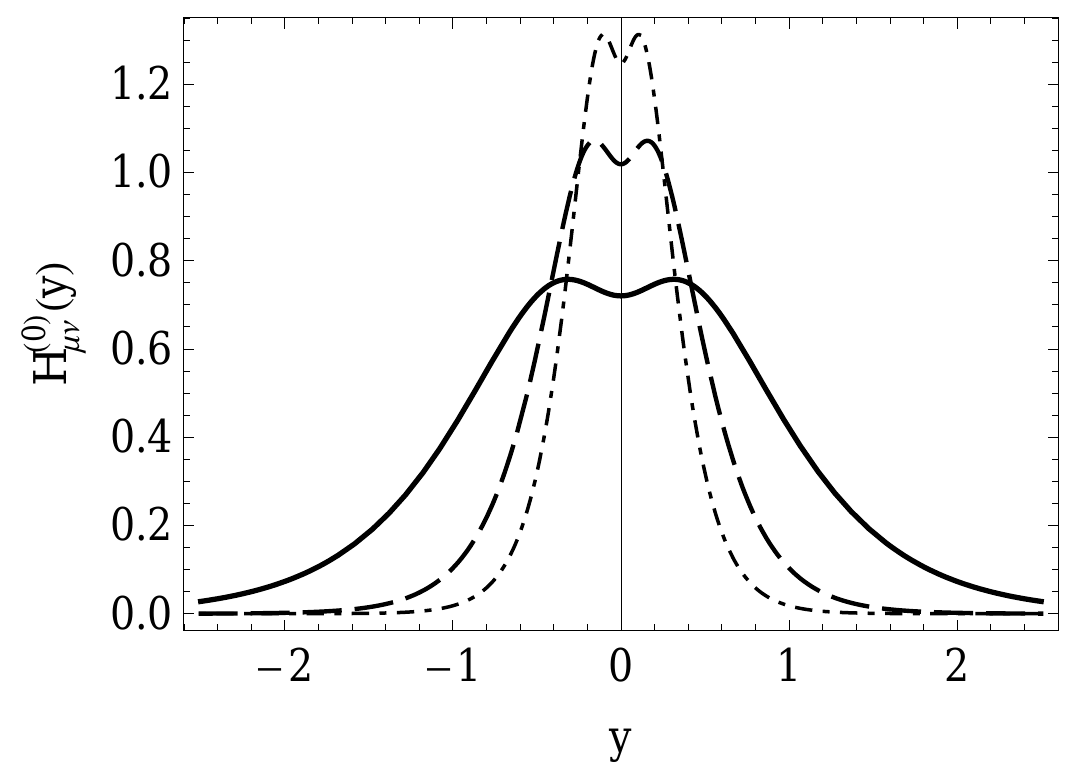}
    \end{center}
    \vspace{-0.9cm}
    \caption{\small{Potential of stability (top panel) and zero-mode (bottom panel) for the model studied in Sec. \ref{modeloA}, represented for $a$ as in Fig. \ref{fig1} and $\psi_0=\varphi_0=10$ (left panel), $\psi_0=-\varphi_0=-10$ (right panel)}. \label{figure21}}
\end{figure}

Figure \ref{figure21} shows the stability potential $\cal U$ and the graviton zero-mode $H_{\mu\nu}^{(0)}$ for the solutions obtained in Sec.~\ref{modeloA}. Note that only the initial conditions that preserve a real mode-zero are considered. In this situation, one verifies that the stability potential might have one of two behaviors depending on the values of $\psi_0$ and $\varphi_0$: either it presents a single potential well at $y=0$ or it presents a potential barrier at $y=0$ with two potential wells. Consequently, the graviton zero-mode will be either a single peak at $y=0$ or develop internal structure, respectively. As the parameter $a$ increases, we also verify that there is a narrowing of the spacial distribution of these functions, while increasing the numerical values at the origin $y=0$. However, the parameter $a$ has no influence in the presence of internal structure, i.e., a variation in $a$ while keeping the remaining parameters constant can not induce or eliminate internal structure, it simply enhances the already existing behavior.

\subsection{Stability of the second model}

\begin{figure}[!htb]
    \begin{center}
        \includegraphics[scale=0.37]{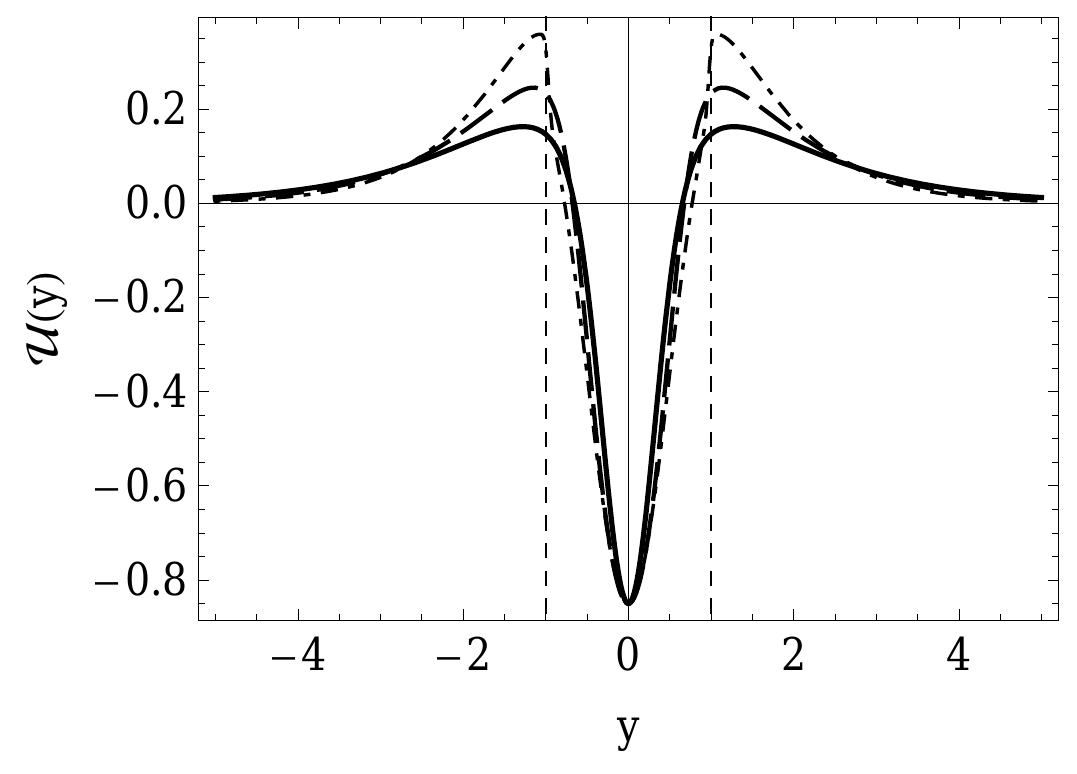}
        \hspace{0.2 cm}
        \includegraphics[scale=0.37]{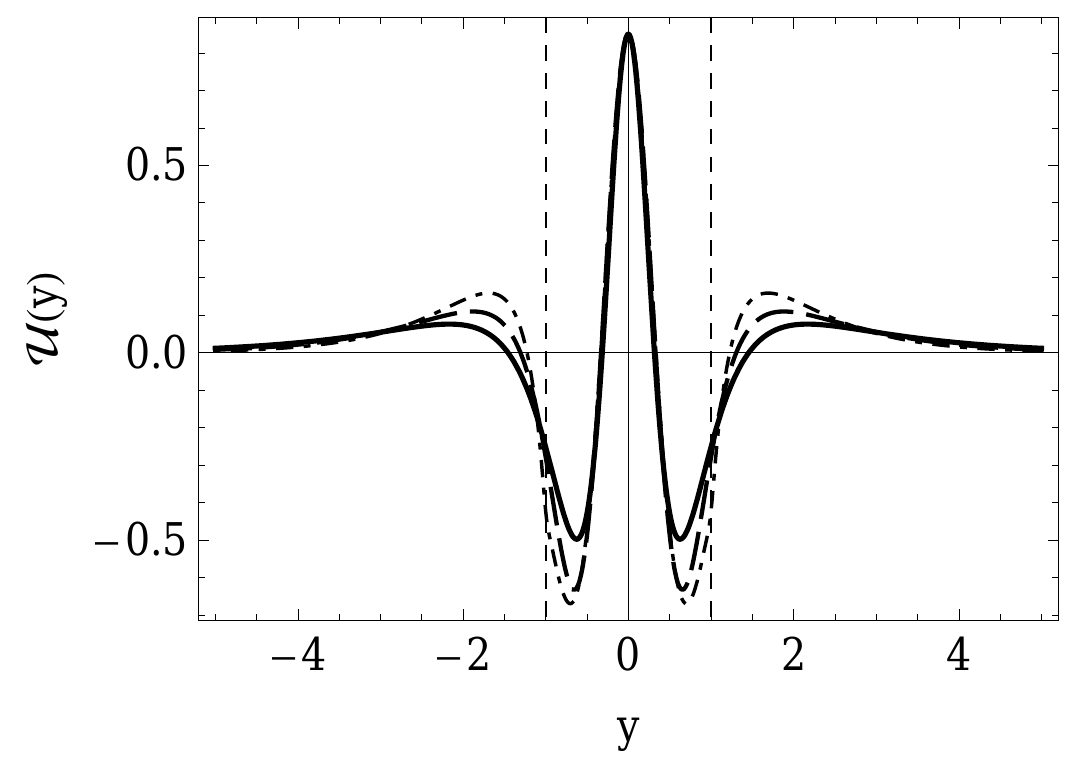}
        \includegraphics[scale=0.37]{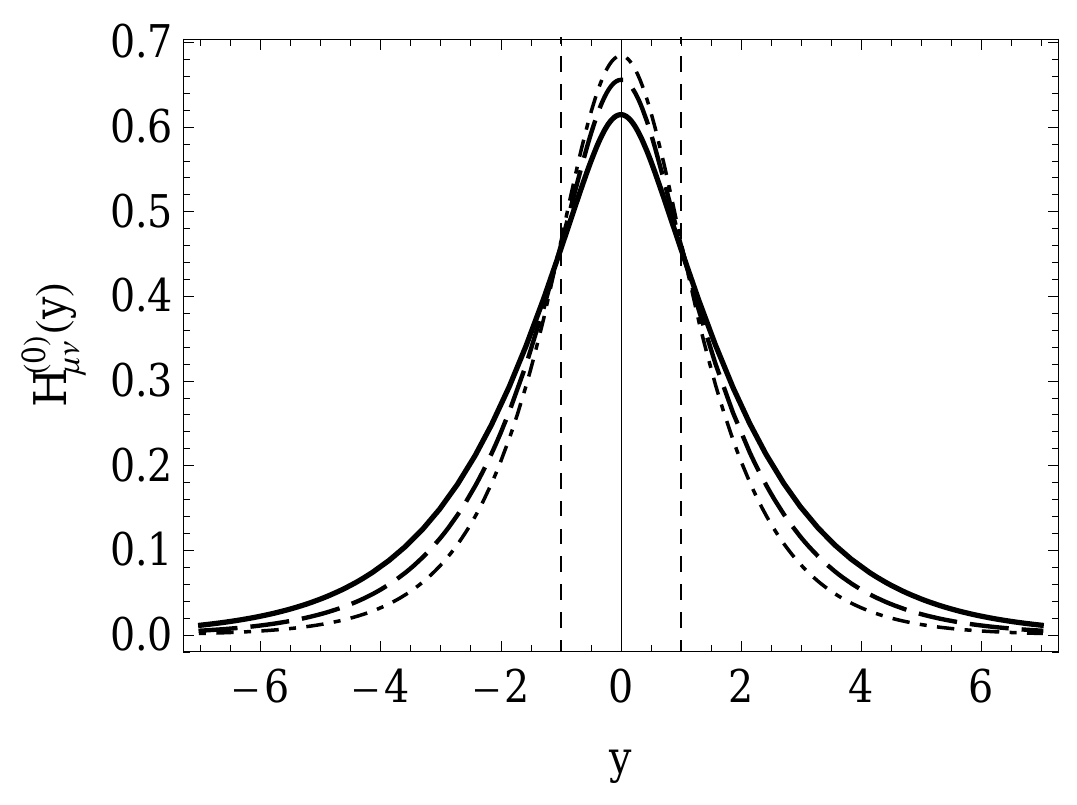}
        \hspace{0.2 cm}
        \includegraphics[scale=0.37]{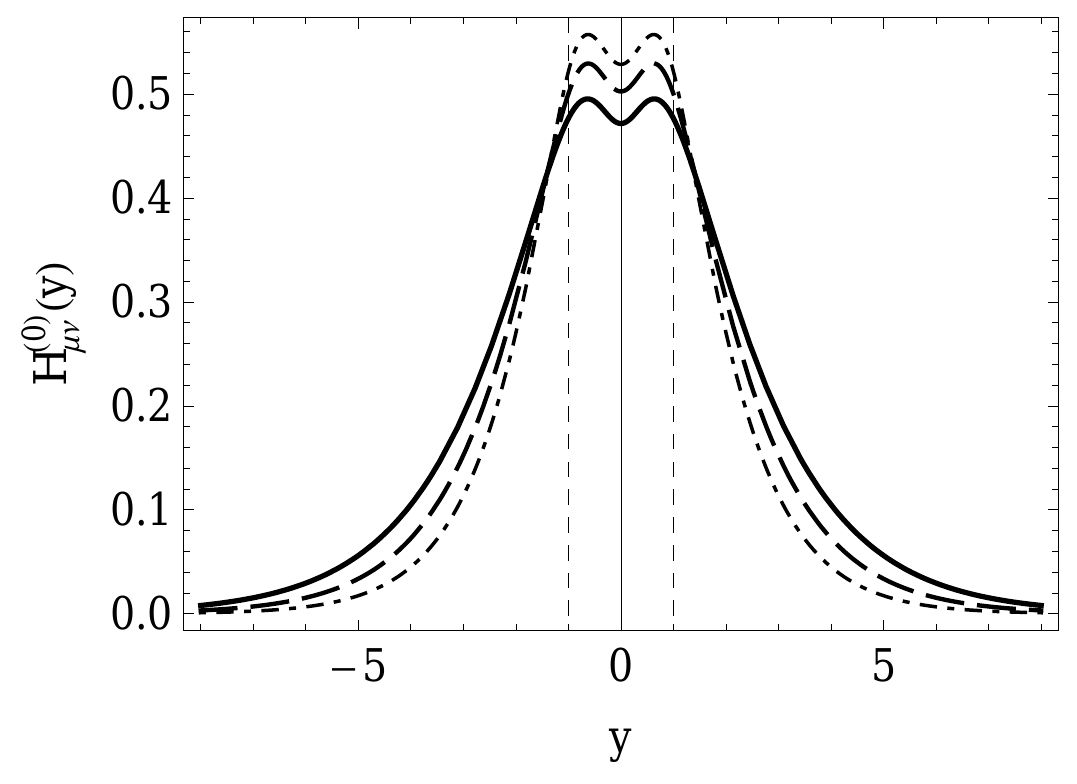}
    \end{center}
    \vspace{-0.9cm}
    \caption{\small{Potential of stability (top panel) and zero-mode (bottom panel) obtained for the model considered in \ref{modeloB} with $\psi_0=\varphi_0=10$ (left panel) and $\psi_0=-\varphi_0=-10$ (right panel) and $n$ as in Fig. \ref{fig6}}. \label{figure22}}
\end{figure}

Fig.~\ref{figure22} shows the stability potential $\cal U$ and the graviton zero-mode $H_{\mu\nu}^{(0)}$ for the solutions obtained in Sec.~\ref{modeloB}. There, we depict the situations for $\psi_0=\varphi_0=10$ and $\psi_0=-\varphi_0=-10$, i.e., conditions that preserve the real nature of the graviton zero-mode. Similarly to the previous case, we verify that the sign of $\psi_0$ controls the qualitative behavior of the stability potential, which can again have a potential-well behavior or a potential-barrier behavior. Consequently, for the first situation the graviton zero-mode will also be a single peak at $y=0$ and for the second situation it develops an internal structure. Furthermore, an increase in the parameter $n$ pushes the spacial distributions of these functions into the region $-1<y<1$, as expected. However, the parameter $n$ does not have any influence on the presence of internal structure, as variations on this parameter leave the central value of the stability potential at $y=0$ unchanged.

\subsection{Stability of the third model}

\begin{figure}[!htb]
    \begin{center}
        \includegraphics[scale=0.37]{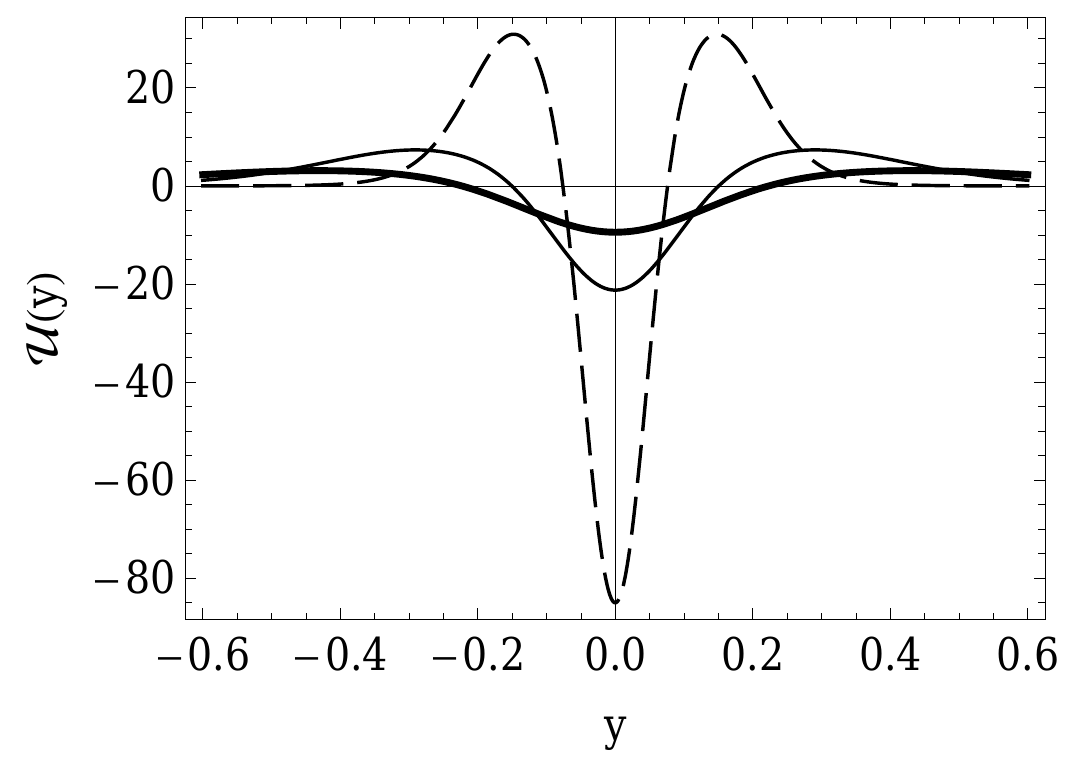}
        \hspace{0.2 cm}
        \includegraphics[scale=0.37]{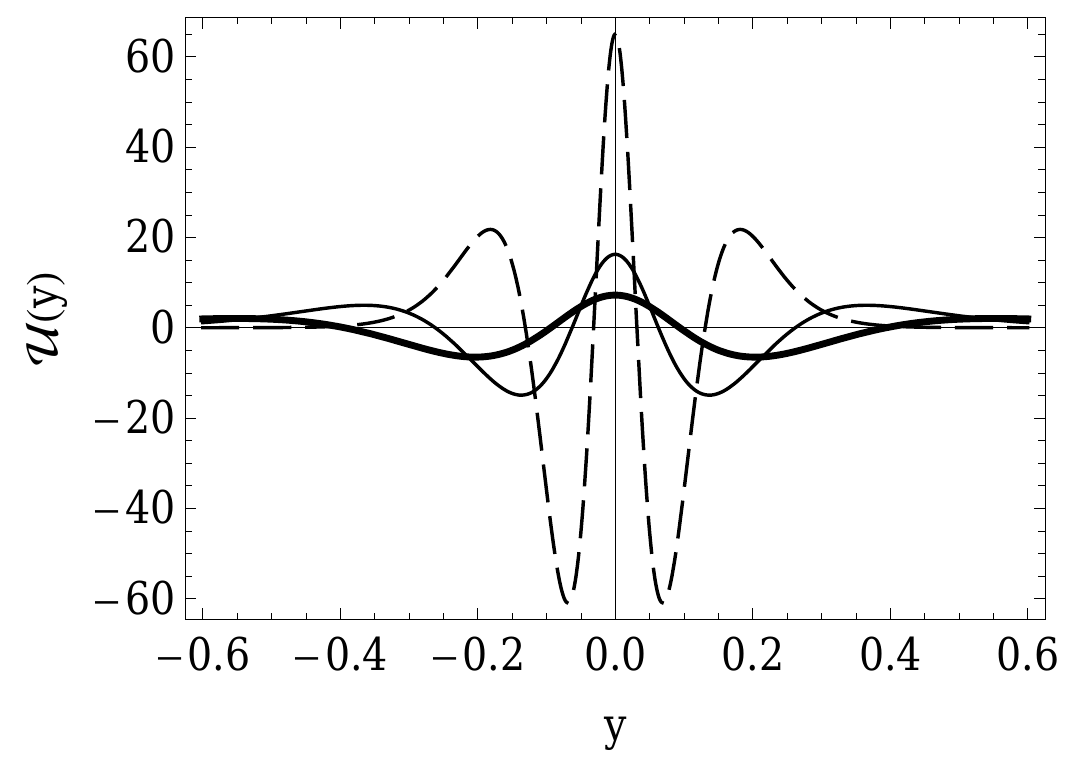}
        \includegraphics[scale=0.37]{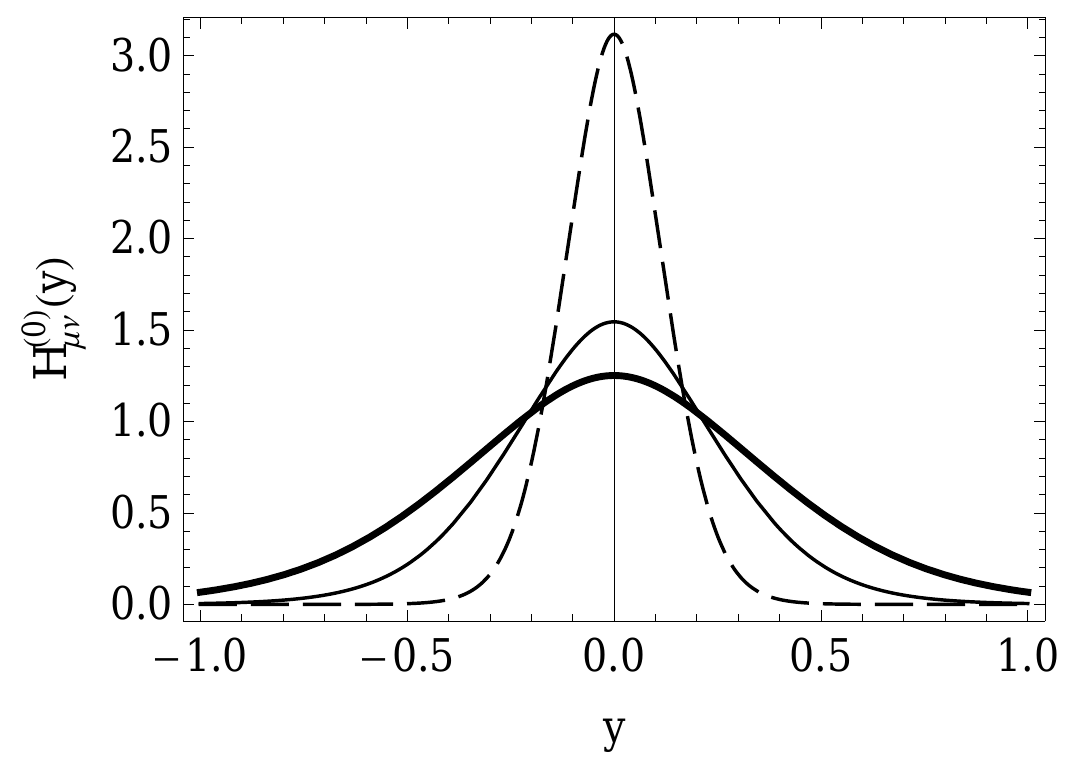}
        \hspace{0.2 cm}
        \includegraphics[scale=0.37]{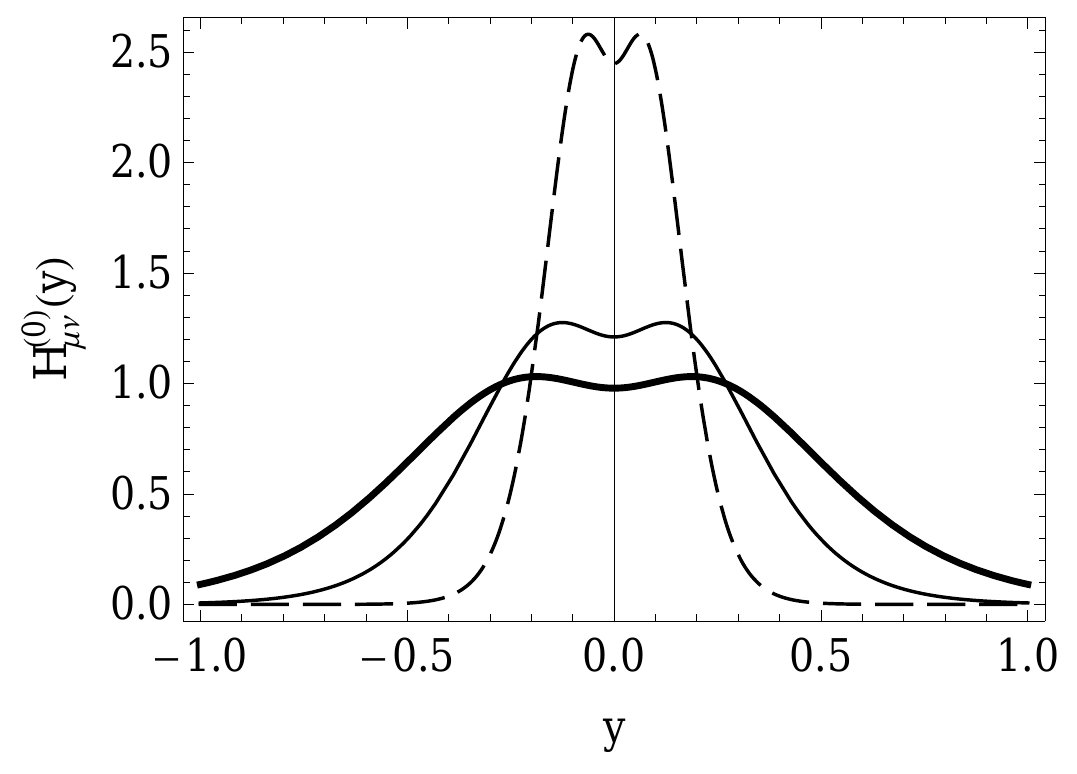}
    \end{center}
    \vspace{-0.9cm}
    \caption{\small{Potential of stability (top panel) and zero mode (bottom panel) for the model studied in Sec. \ref{modeloC}, represented for $\lambda=0.7,\,0.8,\,0.9$ and $\psi_0=\varphi_0=10$ (left panel), $\psi_0=-\varphi_0=-10$ (right panel).}
    \label{figure23}}
\end{figure}

The results for the model where the compactification occurs in the geometry, i.e., the model studied in Sec.~\ref{modeloC}, are shown in Fig. \ref{figure23}. Here we use the same parameter region used before. Note that the quantities are all real only if $\psi_0=\varphi_0=10$ and $\psi_0=-\varphi_0=-10$. Again, the same qualitative behaviors for the stability potential (and consequently for the graviton zero-mode) are obtained, with a clear dependence on the value of $\psi_0$. Furthermore, it is also clear that an increase in the parameter $\lambda$ compactifies the solutions closer to $y=0$, while simultaneously increasing their values at the origin. Similarly to what happens for the model in Sec. \ref{modeloA}, the parameter $\lambda$ does not influence the presence of internal structure, it merely enhances the already present behavior.

\subsection{Stability of the Fourth model}

\begin{figure}[!htb]
    \begin{center}
        \includegraphics[scale=0.37]{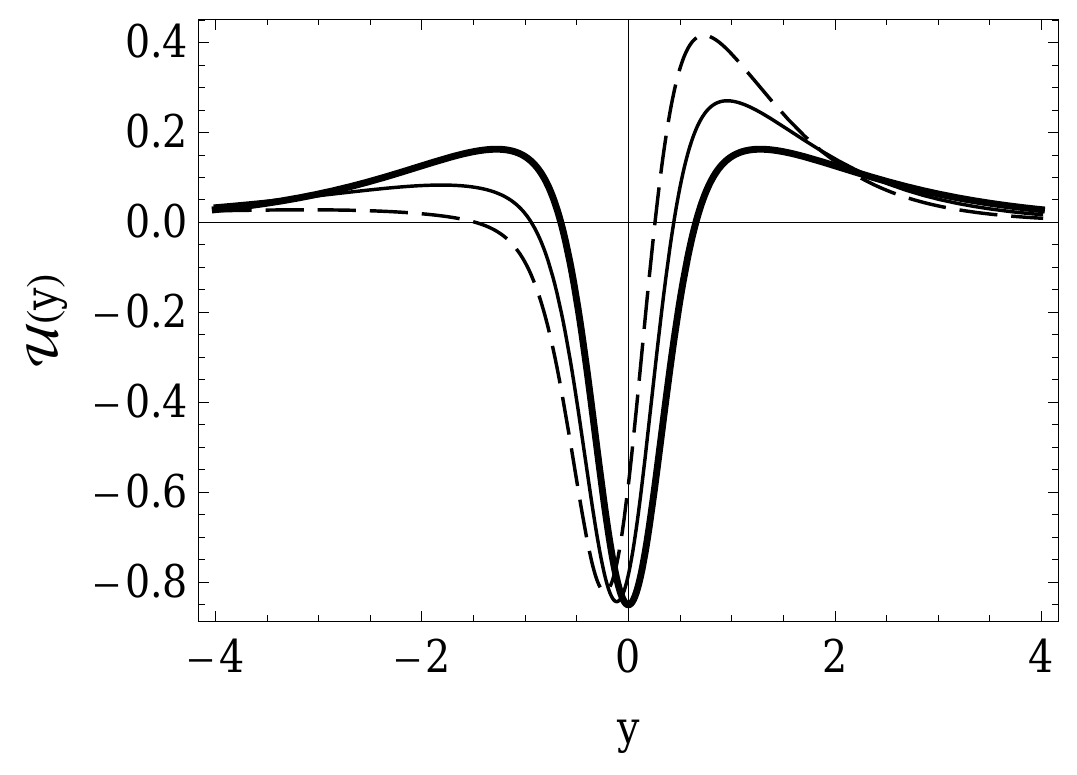}
        \hspace{0.2 cm}
        \includegraphics[scale=0.37]{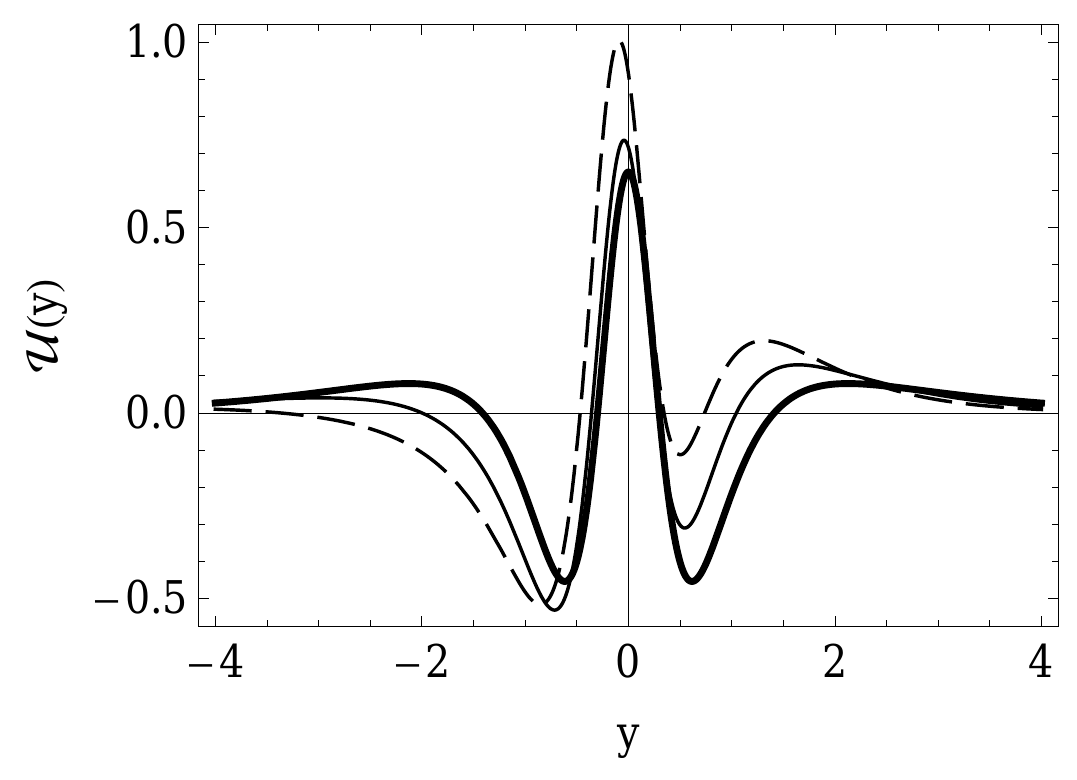}
        \includegraphics[scale=0.37]{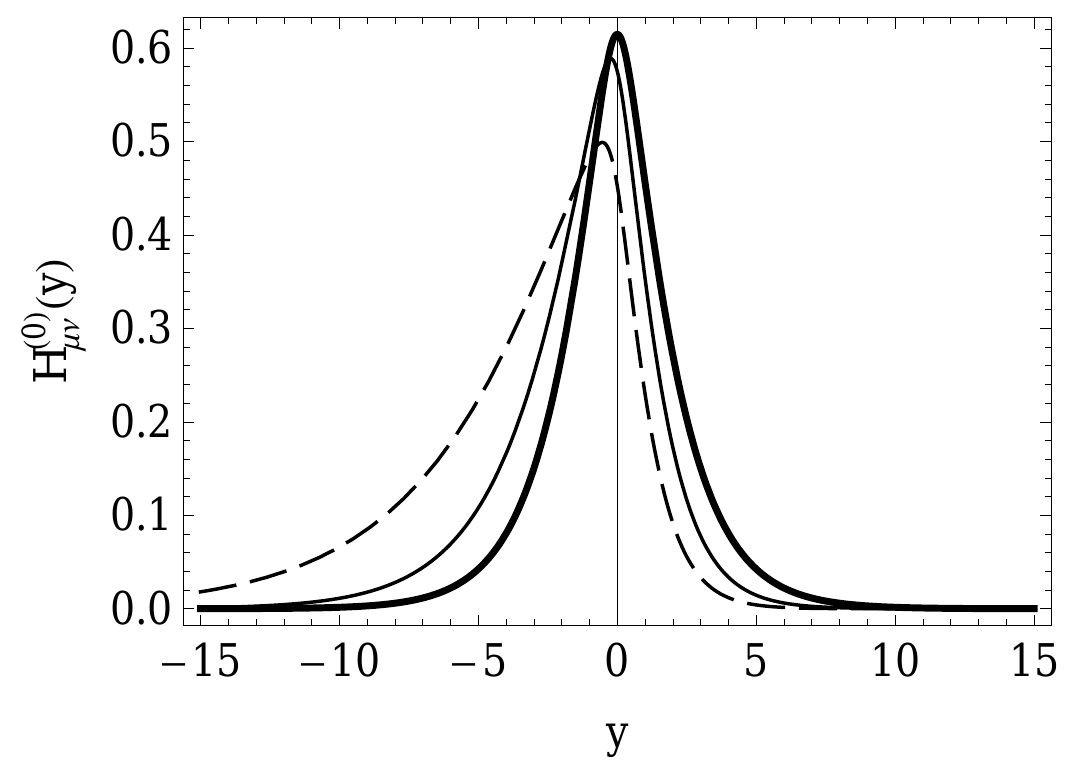}
        \hspace{0.2 cm}
        \includegraphics[scale=0.37]{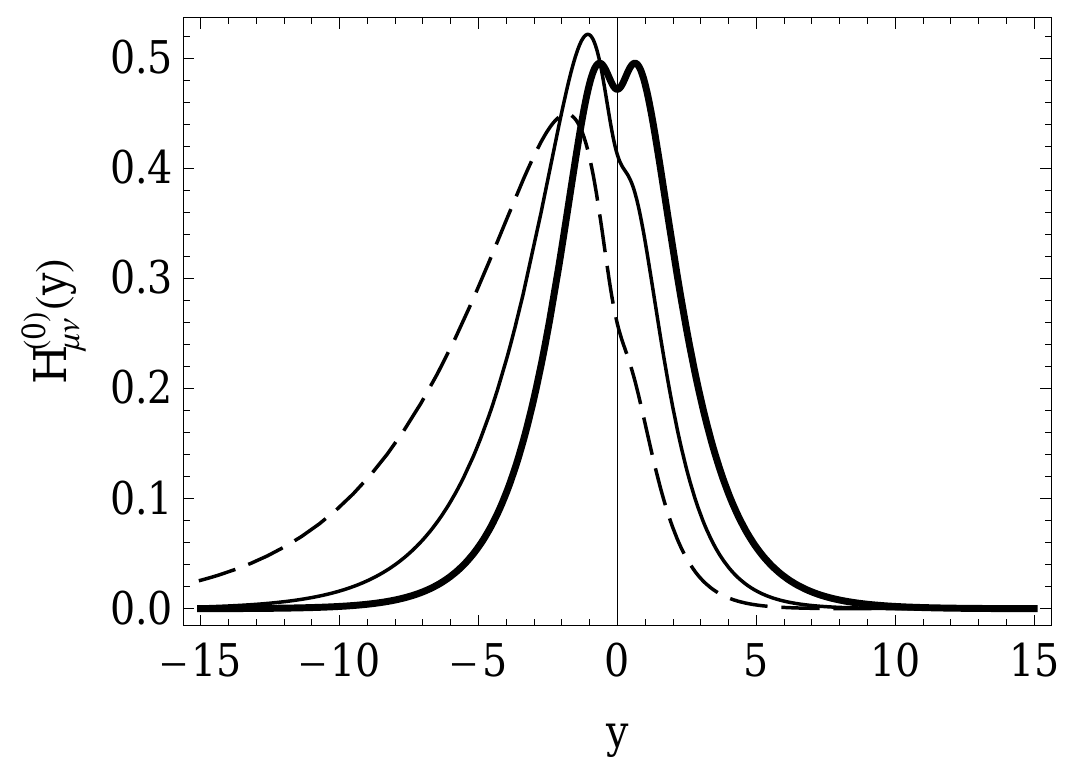}
    \end{center}
    \vspace{-0.9cm}
    \caption{\small{Potential of stability (two top panel) and zero-mode (two bottom panel) obtain for the model of Sec. \ref{modeloD} with $c=0,\,0.2,\,0.4$ and $\psi_0=\varphi_0=10$ (two left panel) and $\psi_0=-\varphi_0=-10$} (two right panel). \label{figure24}}
\end{figure}

Finally, for the model studied in Sec. \ref{modeloD}, we have the behaviors of the stability potential and the graviton zero-mode shown in Fig. \ref{figure24}. Here the potential and the zero-mode were displayed for the initial conditions $\psi_0=\varphi_0=10$, $\psi_0=-\varphi_0=-10$  and $c=0,\,0.2,\,0.4$. For the case $\psi_0>0$ the stability potential is always a single potential well surrounded by two potential barriers. As the value of $c$ increases, the height of one of these barriers decreases, while the other increases. As a consequence, the graviton zero-mode changes from a symmetric single peak at $y=0$ to a peak at some $y_c<0$ that decreases rapidly for $y>y_c$ and slowly for $y<y_c$. In this case, internal structure is never developed. On the other hand, for $\psi_0<0$, we verify that the stability potential corresponds to a potential barrier surrounded by two potential wells. As the value of $c$ increases, one of the potential wells deepens, whereas the other shallows. For $c=0$, the potential barrier present in the stability potential induces an internal structure on the graviton zero-mode. However, as the value of $c$ increases, one of the peaks of the graviton zero mode increases and the other decreases, eventually leading to the disappearance of the local minimum at $y=0$ and a loss of the internal structure.

\section{Comments and conclusions}\label{sec:conclusion}

In this paper, we have investigated the effects on the auxiliary fields of branes described by the scalar-tensor representation of generalized $f(R,T)$ gravity, caused by the presence of compactlike and asymmetric configurations of the source field. We verified that the compactification of the source field of the brane changes not only the behavior of the auxiliary fields but also the structure of the stability potential and zero-mode of the tensor perturbations. To achieve these conclusions, we developed a first-order formalism to establish the relationships between fields and obtain analytical and numerical results for the four different models investigated in this work.

In the first model, studied in Sec. \ref{modeloA}, we verified that the first-order formalism can be used to obtain analytical solutions for the auxiliary field $\psi(y)$. This is a new and interesting result, because it enables an analysis of the asymptotic behavior of this field, regardless of the initial conditions imposed for solving the equations of motion. We verified that the value of the field $\psi$ at the origin is determinant for the behavior of the field $\varphi(y)$ and for the potential $U(y)$. The initial conditions also contribute for the emergence of a inner structure in the zero-mode of the tensor fluctuations. We also verified that the change in the thickness of the solution of the source field interferes with the location of the studied quantities close to the origin, but it does not change the asymptotic values of the fields.

The second model studied in Sec. \ref{modeloB} is particularly interesting in the sense that it seems to induce the appearance of a hybrid profile at the brane, i.e., the brane behaves as a thick brane near the origin and as a thin brane for $|y|>1$. Moreover, for the generalized action used in this work, in the stability analysis we also verified that the stability potential might develop a potential barrier at the origin surrounded by two potential wells, thus inducing a volcano-shaped graviton zero-mode with internal structure.

The third model, studied in Sec. \ref{modeloC}, also presented interesting results. In this case, we verified that with the introduction of a parameter $\lambda$ it was possible to shrink the warp factor, inducing a strong confinement effect in the auxiliary fields $\varphi$ and $\psi$. These effects did not lead to qualitative changes in the auxiliary fields, such as the appearance of a hybrid structure as presented in the second model. However, it showed that both auxiliary fields can be strongly linked to the source field solution. We also studied the stability for this model, and verified that both the stability potential and the graviton zero-mode respond to the effects of shrinking of the warp factor. As in the second model, the potential of stability may also change its shape and induce internal structure in the graviton zero-mode.

We have also studied the issue of asymmetric branes. This was done in Sec. \ref{modeloD}, with the fourth model. The results showed that the brane becomes asymmetric in the presence of nonzero values of the parameter $c$, which controls the asymmetric profile of the system. Moreover, both the $\varphi$ and $\psi$ fields get modified by the presence of the asymmetry. We believe that the modification in $\psi$ is mainly related to the modification of the trace of the stress-energy tensor $T$, which depends on the potential $V(\phi)$ which changes in the presence of $c$. Also, the modification in $\varphi$ is mainly connected to the modification of the warp factor, which is directly modified by the presence of the parameter $c$. In this case, even though the stability potential can develop a potential barrier at the origin, this does not guarantee that the graviton will have a local minimum at $y=0$ due to the asymmetry of the potential.

The internal structure that appears in some braneworld scenarios can change the resonance spectrum and the location of the graviton. In this sense, the profile of the auxiliary fields must somehow interfere with the resonant spectrum of the brane. This is an interesting possibility, that can be studied asking how the auxiliary fields can interfere with the resonance spectrum of the brane in the presence of fermions and gauge fields. We are also interested in other braneworld scenarios, in particular, in mimetic gravity recently considered in Refs. \cite{M0,M1,M2} and also, in the case of braneworld scenarios with bulk fluids parametrized by a nonlinear equation of state \cite{Anto}. Moreover, the Horndeski theory and related generalizations provide interesting possibilities  to study scalar–tensor theories of generalized gravity \cite{Horn} and this adds more motivation to study braneworld issues within the Horndeski scenarios \cite{BH}. These and other related issues are now under investigation and we hope to report on them in the near future.

\begin{acknowledgments}
DB would like to thank CNPq (Brazil), grants No. 404913/2018-0 and No. 303469/2019-6, and Paraiba State research foundation, FAPESQ-PB, grant No. 0015/2019, for partial financial support. JLR was supported by the European Regional Development Fund and the programme Mobilitas Pluss (MOBJD647).
\end{acknowledgments}


\begin{thebibliography}{99}

\bb{RS} L. Randall and R. Sundrum, Phys. Rev. Lett. {\bf83}, 4690 (1999).

\bb{T1}W. D. Goldberger and M. B. Wise, Phys. Rev. Lett. 83, 4922 (1999).

\bb{T2}O. DeWolfe, D. Z. Freedman, S. S. Gubser and A. Karch, Phys. Rev. D 62, 046008 (2000).
\bb{T3} C. Csaki, J. Erlich, T. J. Hollowood and Y. Shirman, Nucl. Phys. B 581, 309-338 (2000).
\bb{C}A. Campos, Phys. Rev. Lett. 88, 141602 (2002)
\bb{B1}D. Bazeia, C. Furtado, A.R. Gomes, JCAP 0402, 002 (2004)
\bb{B2}D. Bazeia, A.R. Gomes, JHEP 0405, 012 (2004)
\bb{Dutra1}A. de Souza Dutra, A. C. Amaro de Faria Jr., M. Hott, Phys. Rev. D 78, 043526 (2008)
\bb{AS0}A. Melfo, N. Pantoja, and A. Skirzewski, Phys. Rev. D 67, 105003 (2003).
\bb{AS1}A. Padilla, Class. Quant. Grav. 22, 681 (2005)
\bb{AS2}A. Padilla, Class. Quant. Grav. 22, 1087 (2005)
\bb{AS3}A. Ahmed, L. Dulny and B. Grzadkowski, Eur. Phys. J. C 74 (2014) 2862
\bb{AS4}D. Bazeia, R. Menezes, and R. da Rocha, Adv. High Energy Phys.  2014, 276729 (2014)
\bb{AS5}A. de Souza Dutra, G. P. de Brito and J. M. Hoff da Silva, EPL 108 11001 (2014).
\bb{AS6}D. Bazeia, M. A. Marques, and R. Menezes, Phys. Rev. D 92, 084058 (2015)
\bb{AS7}R. Menezes and D. C. Moreira, Ann. Phys. 383 (2017) 662
\bb{AS8}D. Bazeia and D. A. Ferreira, Ann. Phys. 411 (2019) 167975
\bb{sken}K. Skenderis and P. K. Townsend,
Phys. Lett. B 468 (1999) 46

\bb{review}V. Dzhunushaliev, V. Folomeev and M. Minamitsuji, Rept. Prog. Phys.  73 (2010) 066901
\bb{Neves:2021dqx}J.~C.~S.~Neves, Phys. Rev. D 104 (2021) 084019

\bibitem{Biggs:2021iqw}
W.~D.~Biggs and J.~E.~Santos, [arXiv:2108.00016 [hep-th]].

\bibitem{Banerjee:2021qei}
S.~Banerjee, U.~Danielsson and S.~Giri, JHEP 09 (2021) 158


\bibitem{Davood:2018} S. Davood Sadatian, S. M. Hosseini,
Adv. High Energy Phys. 2018, 2164764 (2018).

\bibitem{Sui:2020fty}
T.-T.~Sui, W.-D.~Guo, Q.-Y.~Xie and Y.-X.~Liu,
Phys. Rev. D 101 (2020) 055031

\bibitem{Rippl:1995bg}
S.~Rippl, H.~van Elst, R.~K.~Tavakol and D.~Taylor,
Gen. Rel. Grav. 28 (1996) 193

\bibitem{Hwang:2001pu}
J.~C.~Hwang and H.~Noh,
Phys. Lett. B 506 (2001) 13

\bibitem{DeFelice:2010aj}
A.~De Felice and S.~Tsujikawa,
Living Rev. Rel. 13 (2010) 3

\bibitem{RMP}T. P. Sotiriou and V. Faraoni, Rev. Mod. Phys. 82, (2010) 451
\bibitem{Bazeia:2013oha}
D.~Bazeia, R.~Menezes, A.~Y.~Petrov and A.~J.~da Silva,
Phys. Lett. B 726 (2013) 523

\bibitem{Bazeia:2013uva}
D.~Bazeia, A.~S.~Lob\~ao, Jr., R.~Menezes, A.~Y.~Petrov and A.~J.~da Silva,
Phys. Lett. B 729 (2014) 127

\bb{Olmo}D. Bazeia, L. Losano, R. Menezes, Gonzalo J. Olmo, D. Rubiera-Garcia1, Eur. Phys. J. C 75 (2015) 569

\bb{Olmo01}D. Bazeia, A. S. Lobao Jr., L. Losano, R. Menezes, Gonzalo J. Olmo, Phys. Rev. D 91 (2015) 124006

\bb{GB01}S. C. Davies, Phys. Rev. D 72 (2005) 024026
\bb{GB02}H. Maeda, Phys. Rev. D 85 (2012) 124012
\bibitem{Bazeia:2015dna}
D.~Bazeia, A.~Lobao, L.~Losano, R.~Menezes and A.~Y.~Petrov,
Phys. Rev. D 92 (2015) 064010

\bibitem{Bazeia:2015owa}
D.~Bazeia, A.~S.~Lob\~ao and R.~Menezes,
Phys. Lett. B 743 (2015) 98.


\bb{Liu1}Y.-X. Liu, H.-T. Li, Z.-H. Zhao, J.-X. Li, J.-R. Ren, J. High Energy Phys. 0910, 091 (2009)
\bb{Gomes1}C.A.S. Almeida, R. Casana, M.M. Ferreira and A.R. Gomes, Phys. Rev. D 79, 125022 (2009)
\bb{Dutra2} R. A. C. Correa, A. de Souza Dutra, M. B. Hott,
Class. Quant. Grav. 28, 155012 (2011)
\bb{Castro}L. B. Castro, Phys. Rev. D 83, (2011).
\bibitem{Moreira:2021vcf}
A.~R.~P.~Moreira, J.~E.~G.~Silva and C.~A.~S.~Almeida,
Eur. Phys. J. C 81 (2021) 298

\bibitem{Yang:2012hu}
J.~Yang, Y.~L.~Li, Y.~Zhong and Y.~Li,
Phys. Rev. D 85 (2012) 084033

\bibitem{Yu:2015wma}
H.~Yu, Y.~Zhong, B.~M.~Gu and Y.~X.~Liu,
Eur. Phys. J. C 76 (2016) 195

\bibitem{Tan:2020sys}
Q.~Tan, W.~D.~Guo, Y.~P.~Zhang and Y.~X.~Liu,
Eur. Phys. J. C 81 (2021) 373

\bibitem{Rosa:2021tei}
J.~L.~Rosa, M.~A.~Marques, D.~Bazeia and F.~S.~N.~Lobo,
Eur. Phys. J. C 81 (2021) 981

\bibitem{Rosa:2021teg}
J.~L.~Rosa, Phys. Rev. D 103 (2021) 104069

\bb{Lobo1}T. Harko, F.S.N. Lobo, S. Nojiri, S.D. Odintsov,
Phys. Rev. D 84, 024020 (2011).

\bibitem{Rosa:2021ld}
J.~L.~Rosa, D.~Bazeia and A.~Lobão S. Jr.,
[arXiv:2111.08089 [hep-th]].


\bibitem{Adam:2007ag}
C.~Adam, N.~Grandi, J.~Sanchez-Guillen and A.~Wereszczynski,
J. Phys. A 41 (2008) 212004
[erratum: J. Phys. A \textbf{42} (2009), 159801]

\bibitem{Bazeia:2008zx}
D.~Bazeia, A.~R.~Gomes, L.~Losano and R.~Menezes,
Phys. Lett. B 671 (2009) 402

\bibitem{Bazeia:2008tj}
D.~Bazeia, L.~Losano and R.~Menezes,
Phys. Lett. B 668 (2008) 246

\bb{C01}A. Ito, Y. Sakakihara, and J. Soda, Phys. Rev. D 100, 063531 (2019).
\bb{C02}A. Iyonaga, K. Takahashi, and T. Kobayashi, JCAP 07, 004 (2020).
\bb{C03}J. Quintin and D. Yoshida, JCAP 02, 016 (2020).
\bibitem{Andrade:2018afh}
I.~Andrade, M.~A.~Marques and R.~Menezes,
Nucl. Phys. B 942 (2019) 188

\bibitem{Bazeia:2021jok}
D.~Bazeia, D.~A.~Ferreira and M.~A.~Marques,
Eur. Phys. J. C 81 (2021) 619


\bibitem{Bazeia:2014hja}
D.~Bazeia, L.~Losano, M.~A.~Marques and R.~Menezes,
Phys. Lett. B 736 (2014) 515


\bibitem{Bazeia:2015eta}
D.~Bazeia and D.~C.~Moreira,
Phys. Lett. B 748 (2015) 79

\bb{M0}Y. Zhong, Y. Zhong, Y.-P. Zhang and Y.-X. Liu, Eur. Phys. J. C 78 (2018) 45
\bb{M1}J. Chen, W.-D. Guo and Y.-X. Liu, Eur. Phys. J. C 81 (2021) 709
\bb{M2}T.-T. Sui, Y.-P. Zhang, B.-M. Gu and Y.-X. Liu, Eur. Phys. J. C 81 (2021) 980

\bb{Anto}I. Antoniadis, S. Cotsakis and I. Klaoudatou,  Eur. Phys. J. C 81 (2021) 771
\bb{Horn}T. Kobayashi, Rept. Prog. Phys. 82 (2019) 086901
\bb{BH}Q.-M. Fu, H. Yu, L. Zhao, and Y.-X. Liu, Phys. Rev. D 100, 124057 (2019)

\end{thebibliography}
\end{document}